\documentclass[onecolumn]{aastex6}

\begin{document}

\title{The Radial Acceleration Relation (RAR): the crucial cases of  Dwarf Discs and of Low Surface Brightness galaxies}

\author{C. Di Paolo\altaffilmark{1}}\author{P. Salucci\altaffilmark{1, 2}}\author{J. P.  Fontaine\altaffilmark{3}}


\altaffiltext{1}{
SISSA/ISAS, Via Bonomea 265, 34136 Trieste, Italy}
\altaffiltext{2}{
INFN, Sezione di Trieste, Via Valerio 2, 34127 Trieste, Italy}
\altaffiltext{3}{
GSSI, Viale Francesco Crispi 7, 67100  L'Aquila, Italy}

\begin{abstract}

\noindent
McGaugh et al. (2016) have found, in a large sample of disc systems, a tight
nonlinear relationship between the total radial accelerations $g$ and their components $g_b$  arisen from the distribution of the baryonic matter 
~\citep{McGaugh_2016}. Here, we investigate the existence of such relation in Dwarf Disc Spirals and Low Surface Brightness
galaxies on
the basis of  ~\cite{Karukes_2017} and ~\cite{DiPaolo_2018}. We have accurate
mass profiles for 36 Dwarf Disc Spirals and 72 LSB galaxies. These galaxies have accelerations that cover the McGaugh range
but also reach out to one order of magnitude
below the smallest accelerations present in McGaugh et al. (2016) and span different Hubble Types. 
We found, in our samples, that the $g$ vs $g_b$ relation
has a very different profile and also other intrinsic novel properties,  among those,   the dependence on a second variable: the
galactic radius, normalised to the optical radius $R_{opt}$, at which the two accelerations are measured. 
We show that the new far than trivial $g$ vs $(g_b, r/R_{opt})$ relationship is nothing else than a direct consequence
of the complex, but coordinated mass distributions of the baryons and the dark matter (DM) in disc systems. 
Our analysis shows that the McGaugh et al. (2016) relation is a limiting case of a new universal relation
that  can be very well framed in  the standard "DM halo in the
Newtonian Gravity" paradigm.

\end{abstract}


\maketitle

\vspace{2em}

\section{Introduction} \label{Introduction}
\noindent
 A recent study ~\citep{McGaugh_2016}, hereafter referred to as McG+16, claims an empirical discovery that would challenge 
the idea of dark matter halos surrounding galaxies, or, at least, it would revolutionise our knowledge about the nature of the
huge mass discrepancy therein. The standard paradigm relies on collisionless non luminous particles  constituting about
$25 \%$ of the mass energy of the Universe and playing a crucial role on  the birth and  the evolution of its structures. 

The relation, in rotating systems,  between the galaxy gravitational potential $\Phi_{tot}$
and the radial acceleration $g(r)$ of a point mass at distance $r$ is
\begin{eqnarray}
\label{gravity}
g(r) = \frac{V^2(r)}{r} = \left| -\frac{d \, \Phi_{tot}(r)}{d \, r} \right|   \quad ,
\end{eqnarray}   
with $V(r)$ the circular velocity. The baryonic component of the radial acceleration is given by:
\begin{eqnarray}
\label{baryon_gravity}
g_b(r) = \frac{V^2_b (r)}{r} = \left|-\frac{d \, \Phi_{b}(r)}{d \, r} \right|   \quad ,
\end{eqnarray}   
where
\begin{eqnarray}
\label{Baryonic_velocity}
V_b^2 (r) = V_d^2(r) + V_{HI}^2(r)+V_{bu}^2(r)     \quad   
\end{eqnarray}   
is the baryonic contribution to the circular velocity.
In Eq. \ref{Baryonic_velocity}, the velocities
$V_i = |-r \, d\Phi_i(r)/dr|^{1/2}$   
are the solutions of the separated Poisson equations:
$\nabla ^2 \Phi_i(r) = 4\pi G\rho _i$.  
$\rho_i$ is equal to the stellar disc, the HI disc and the bulge mass densities and $\Phi _i$ are the gravitational potentials of the $i$-components. Obviously we have:
\begin{eqnarray}
\label{DM_gravity}
g_h(r)=g(r)-g_b(r) \quad , 
\end{eqnarray}   
where $g_h$ refers to the dark matter contribution to the radial acceleration $g$.
\begin{figure*}[!t]
\begin{center} 
\includegraphics[width=0.7\textwidth,angle=0,clip=true]{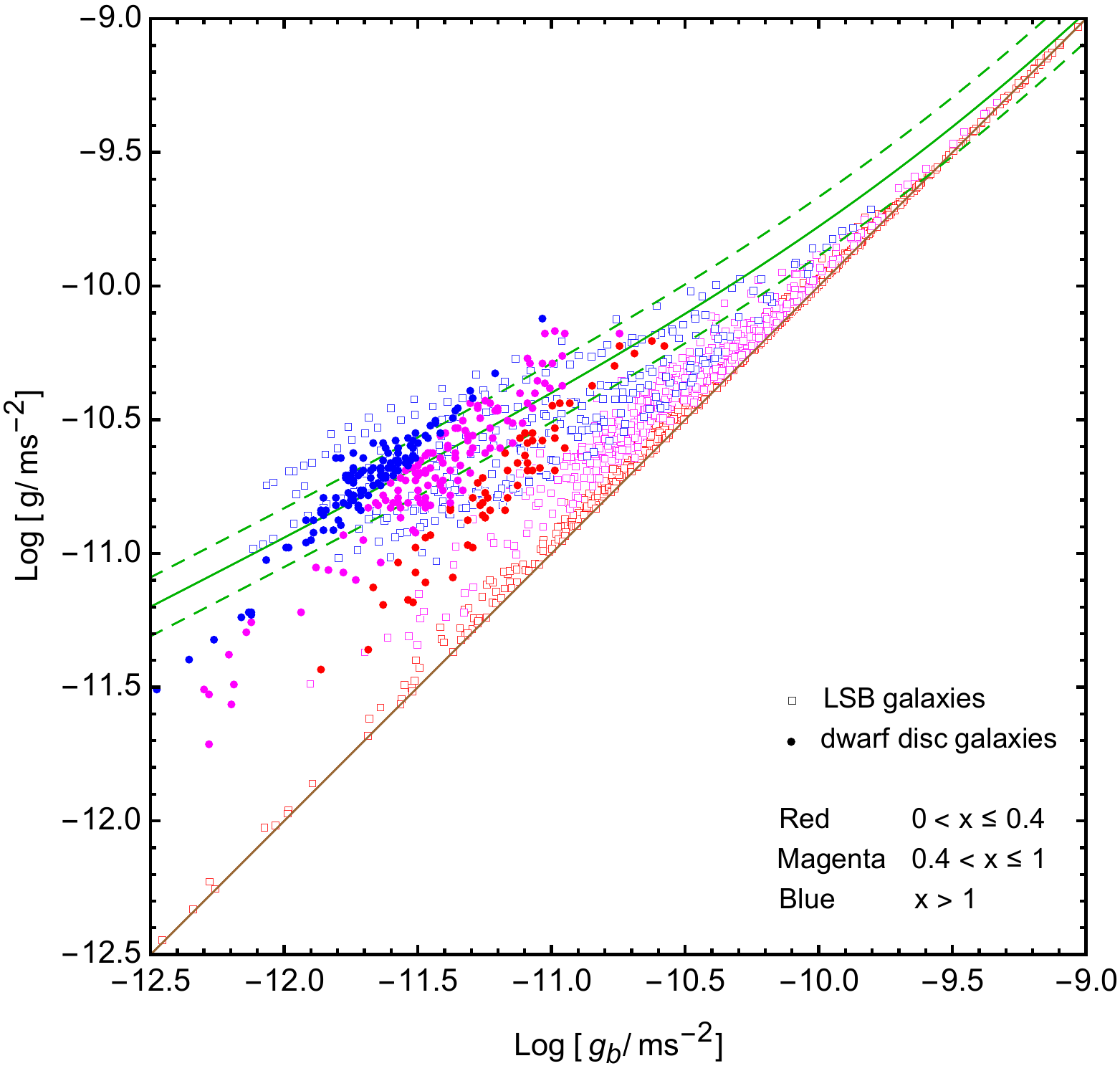}
\caption{Relationship between the total acceleration $g$ and its baryonic component $g_b$. $x=r/R_{opt}$. 
{\it Red}, {\it magenta} and {\it blue points} correspond to radial bins with increasing distance from the  galactic center (see {\it legend}). 
Also shown:
the ~\cite{McGaugh_2016} relationship ({\it solid green line}) with its $1\, \sigma$ errorbars of 0.11 dex ({\it dashed
green lines}); the Newtonian relationship $Log \, g = Log\, g_b$ ({\it brown line}). See also Fig. \ref{g_gb_extension} in Appendix \ref{The extended $g - g_b$ plane}, for LSBs data with 
very low values of $Log \,g$ and $Log \,g_b$.
}  
\label{g_gb_dd_LSB_2D}
\end{center}
\end{figure*}

\noindent
McG+16 investigated 153 galaxies across a wide range of Hubble types and luminosities with new high-quality data from
the Spitzer Photometry and Accurate Rotation Curves (SPARC) database. The analysis includes (see ~\cite{Lelli_2016} for
details):

{\it i)} near-infrared ($3.6 \mu m$) observations that trace the distribution of stellar masses under the assumption of 0.5 $M_{\odot}/L_{\odot}$  for the value of the stellar mass to light ratio in this band;

{\it ii)} the 21 cm observations that trace the distribution of the atomic gas and the velocity fields.

\noindent
They found that the radial acceleration $g(r)$ shows an anomalous feature: it correlates at any radius
and in any object, with its component generated from the baryonic matter $g_b(r)$ in a way that it is : 

{\it i)} very different from the $g = g_b$ relationship expected in the Newtonian case with the presence of the only baryonic matter;

{\it ii)} claimed of difficult understanding in the standard Newtonian + dark matter halos scenario.

\noindent
In detail, the McGaugh relationship (see Fig 1 and Fig 3 in McG+16) relies on 
153 objects for a number of 2693 independent circular velocity measurements. Each of them yields the pairs ($g_b$,$ \,g$),
well fitted by:
\begin{eqnarray}
\label{g_McG+16}
Log \, g(r) = Log \, \left( \frac{g_b(r)}{1-exp \left(-\sqrt{\frac{g_b(r)}{\tilde{g}}} \right)}   \right)\quad ,
\end{eqnarray}   
with $\tilde{g}= 1.2 \times 10^{-10} \, m s^{-2}$.
 At high accelerations, $g \gg \tilde{g}$, Eq. \ref{g_McG+16} converges to the Newtonian relation
$g = g_b$; while, at  lower accelerations, $g < \tilde{g}$, Eq. \ref{g_McG+16} strongly deviates from the latter
 ~\citep{McGaugh_2016, McGaugh_2018}. 

A recent investigation of the McG+16 relationship has been  performed by  ~\cite{Salucci_2018, Salucci_a_2018} (hereafter S18) in  three very large samples of {\it normal spirals} by
exploiting  three specifically devised 
methods of  deriving $g_h$ (as shown in Eq. \ref{DM_gravity}). In these works, the stellar mass distribution is estimated kinematically, by means of the mass modelling of the rotation curves, rather than being estimated from 
spectrophotomery as in McG+16. The outcome is a $g(g_b)$ relationship,  with a r.m.s. of 0.15 dex and
with a quite  small systematical difference of 0.2 dex from Eq.  \ref{g_McG+16} ~\citep{Salucci_a_2018}.
These results, totally framed in the DM scenario and  obtained by means of  novel methods of mass modelling, confirm the McG+16 relationship in normal Spirals.

Recently,  ~\cite{Karukes_2017} and  ~\cite{DiPaolo_2018}  
have obtained the radial distribution of the total, baryonic and dark matter for 36 dwarf spirals, yielding 315 acceleration measurements,  and  72 Low Surface Brightness (LSB)
galaxies, yielding 1601 acceleration measurements (see also Appendix \ref{The extended $g - g_b$ plane} for further details).
These accelerations occupy  a region  in the $g - g_b$ plane (see Fig. \ref{g_gb_dd_LSB_2D}) compatible
with that covered by previous works, but  that, in addition: 

a) reaches smaller values along the vertical axis, considering our smallest value of  
$Log \, g/ m s^{-2} \simeq -12.5$ (-14.5, see Appendix  \ref{The extended $g - g_b$ plane}) and the McG+16 smallest unbinned value of $Log \, g/ m s^{-2} \simeq -11.4$;

b) pertains to different Hubble Types 
than the bulk of the objects in McG+16; it is worth to specifying that the sample of McG+16 (153 rotating objects) has  dwarf and LSB discs alongside with a large number of normal Spirals. In our work, we have only dwarf discs (here called DD) and LSB galaxies. 
\\
\\
A very important element of our analysis is the baryonic fraction $f_b(r)$, which varies in galaxies of different dimensions and Hubble Types. It
will pivotal to frame our data and those of McG+16 and S18 within the standard "DM halo in the Newtonian Gravity" paradigm. Moreover, we will understand why the McG+16 relation is only a limit of a more complex universal relation.
\\
\\
Let us define the distribution of stars in disc galaxies, by means of their surface brightness, which is almost always given, in disc systems, by 
$ \mu(r)= \mu(0)+ 1.086 \,r/R_d$ ~\citep{Freeman_1970}, where 
$R_d$ is the exponential disc scale length ($\mu(0)$ is variable object by object).
In this work, the accelerations are in $m/s^2$ and the distances in kpc. The optical radius $R_{opt}$ is defined as the radius encompassing 83\% of the total luminosity; $R_{opt} = 3.2\,R_d$. The optical velocity $V_{opt}$ is the circular velocity measured at $R_{opt}$.
Let us notice that in this paper, we will use alternatively the quantity $x$ and $r/R_{opt} \equiv x$. In addition, our system 
of coordinates is $r, \,\varphi, \, z$.
\\
\\
The work is organised as follows: in section \ref{The dd and LSB samples}, we will describe the dwarf discs and LSBs samples; in section \ref{The URC method}, we briefly describe the Universal Rotation Curve method used in our analysis; in section \ref{Test for $g$ and $g_b$}, we build the $g$ vs $g_b$ relation followed, in section \ref{The true universality of the $g$ vs $g_b$ relationship}, by a 3D analysis that involves the baryonic fraction $f_b(r)$ and the additional variable $x$. Finally, in section \ref{Conclusion} we report the  consequences of our results.
\begin{figure}[!t]
\begin{center} 
\includegraphics[width=0.70\textwidth,angle=0,clip=true]{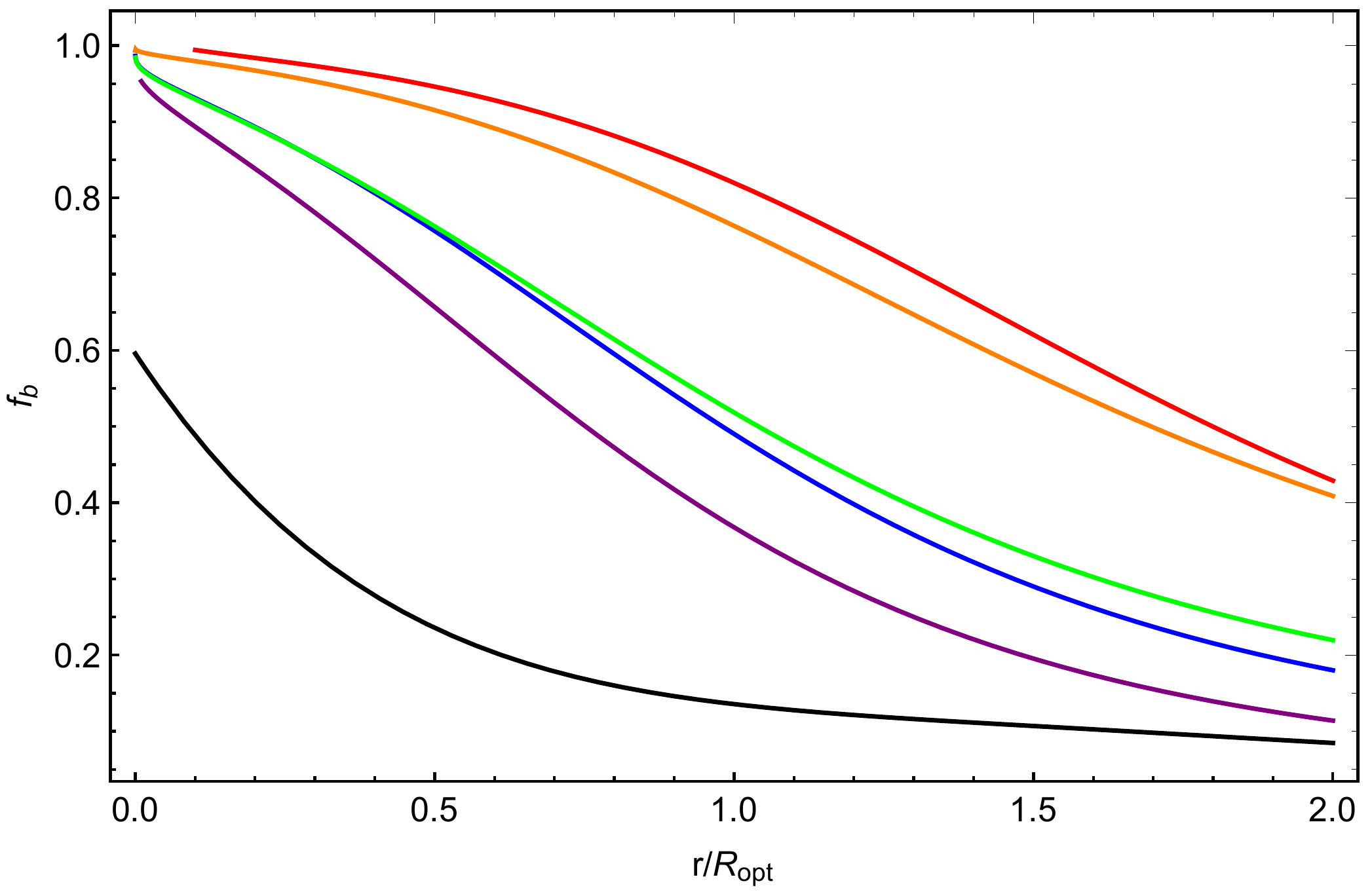}
\caption{Baryonic fraction as function of $r/R_{opt}$, derived by the URCs of DD ({\it black line}, with $\langle V_{opt} \rangle  = 40 \, km/s$)~\citep{Karukes_2017} and of LSBs ({\it purple, blue, green, orange} and {\it red}, with: $ \langle V_{opt} \rangle = 43, \,73, \, 101, \, 141, \, 206 \, km/s$)~\citep{DiPaolo_2018}. For the uncertainties on the $f_b(x)$ see text and Appendix \ref{Small errors on $f_b$: small effect on the $g$ - $g_b$ plane}.}  
\label{fb_plot_LSB}
\end{center}
\end{figure}

\section{The dd and LSB samples}\label{The dd and LSB samples}
\noindent
The sample of dwarf discs ~\citep{Karukes_2017}  that we use in this work is drawn from the Local
Volume catalog ~\citep{Karachentsev_2013}.  The
faintest objects  are  3 magnitudes fainter with respect to the sample of spirals of McG+16 and S18. These 
galaxies explore quite smaller mass scales than the normal Spirals.
The criteria adopted to select the objects are described in ~\citep{Karukes_2017}.
In detail, the sample consists of 36 galaxies (two among them are in common with the LSB sample) whose structural
properties span the intervals: $-19.9  \lesssim M_K  \lesssim  -14.2 \,$, $0.18 \, kpc   \lesssim R_d  \lesssim  1.63 \, kpc
\,$, $17 \,  km/s   \lesssim V_{opt}   \lesssim   61 \, km/s$. All galaxies are 
bulgeless disc systems in which the rotation, corrected for the pressure support, totally balances the gravitational force.

The sample of LSBs consists of 72 disc galaxies.
They are objects which emit an amount of light per area much smaller than normal spirals ~\citep{deBlock_2000, 
McGaugh_1994, Bothun_1997} and don't lay on the $L \propto R_d^2$ relationship of the latter.
The sample of rotation curves is selected from literature (Erkurt et al. in preparation)\footnote{In Appendix \ref{The LSB sample} we provide  the references for the RC data and other galactic properties (see Tab. \ref{LSB_sample_Tab}).
}  and characterised by objects whose optical velocities
$V_{opt}$  span from $\sim 24$ km/s to $\sim 300$ km/s. 

For both  DD and  LSB samples, the available photometry and kinematics are of sufficient quality to allow us  to obtain a proper mass modeling, by means of the technique of the Universal Rotation Curve (URC) ~\citep{Persic_1996}. 

\section{The  mass distribution in disc systems by exploiting  the URC }\label{The URC method}
\noindent
The URC compacts  the structural properties  
of rotating systems ~\citep{Persic_1996, Salucci_2007}. As starting point, all galaxies of a given sample are binned in different groups/families according to their $V_{opt}$ (in the case of our samples) and then co-added in terms of $x\equiv r/R_{opt}$, their radial normalised  coordinate. 
Galaxies inside a certain limited range of $V_{opt}$ have, approximately, all the same baryonic and DM distribution, once they are expressed in normalised radial coordinate $x$. For the present samples, the DD
galaxies are grouped in a single bin ~\citep{Karukes_2017} and the LSB
galaxies are grouped in five bins (according to their increasing $V_{opt}$) ~\citep{DiPaolo_2018}. 
\begin{figure*}[!t]
\begin{center} 
\includegraphics[width=0.74\textwidth,angle=0,clip=true]{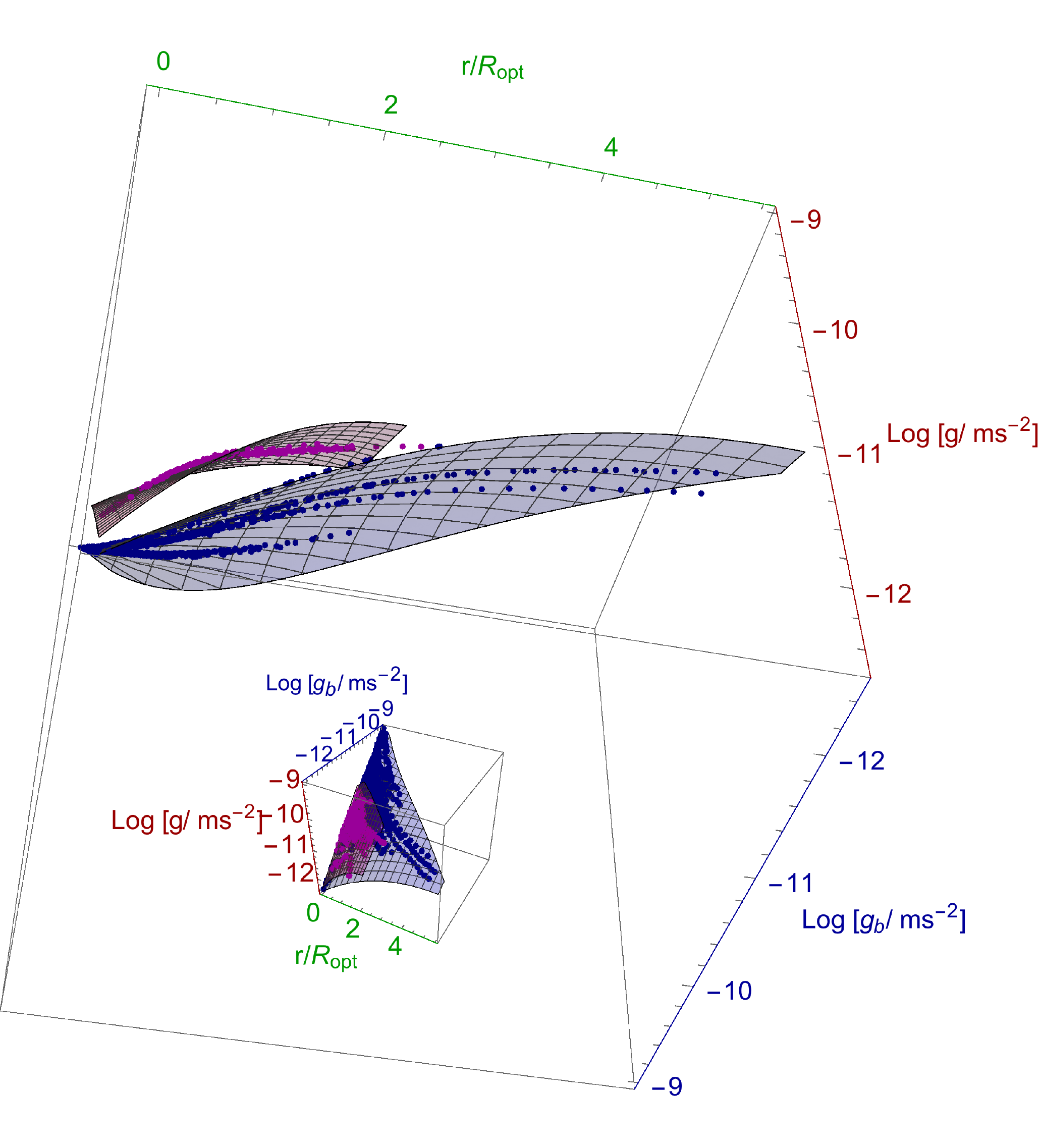}
\caption{Relation among total acceleration $g$, baryonic acceleration $g_b$ and normalised radii $r/R_{opt}$. The {\it magenta} and {\it blue points} refer to DD and LSB galaxies data respectively. The {\it surfaces} are the results from the best fit models.}  
\label{3D_g_gb_x_a}
\end{center}
\end{figure*}

The URC model is based on an exponential disc ~\citep{Freeman_1970} for the stellar component and the Burkert density profile ~\citep{Burkert_1995} for the dark matter halo (preferred in discs systems, see ~\citep{Salucci_2000a, Karukes_2017, deBlok_2002}). For the disc component, the Tonini et al. HI disc ~\citep{Tonini_2006, Evoli_2011} is considered in DD galaxies and a bulge component ~\citep{Yegorova_2007} is taken into account for the largest LSB galaxies ~\citep{Das_2013} . Let us notice that, for LSBs, 
the gas contribution to the circular velocity can be considered negligible in view of the aim of this paper. See Appendix \ref{The gas effect on the $g$-$g_b$ plane}.  

We fit with the URC the co-added rotation curves for each of the 1 + 5 families.  
 This provides us with 
$V_{URC}(r/R_{opt}, V_{opt})$  and $V_{URC, \, b}(r/R_{opt}, V_{opt})$, i.e. the circular velocity and its baryonic component 
(see Appendix \ref{TheURC method} for further details about the URC method).
\begin{figure}[!t]
\begin{center} 
\includegraphics[width=0.9\textwidth,angle=0,clip=true]{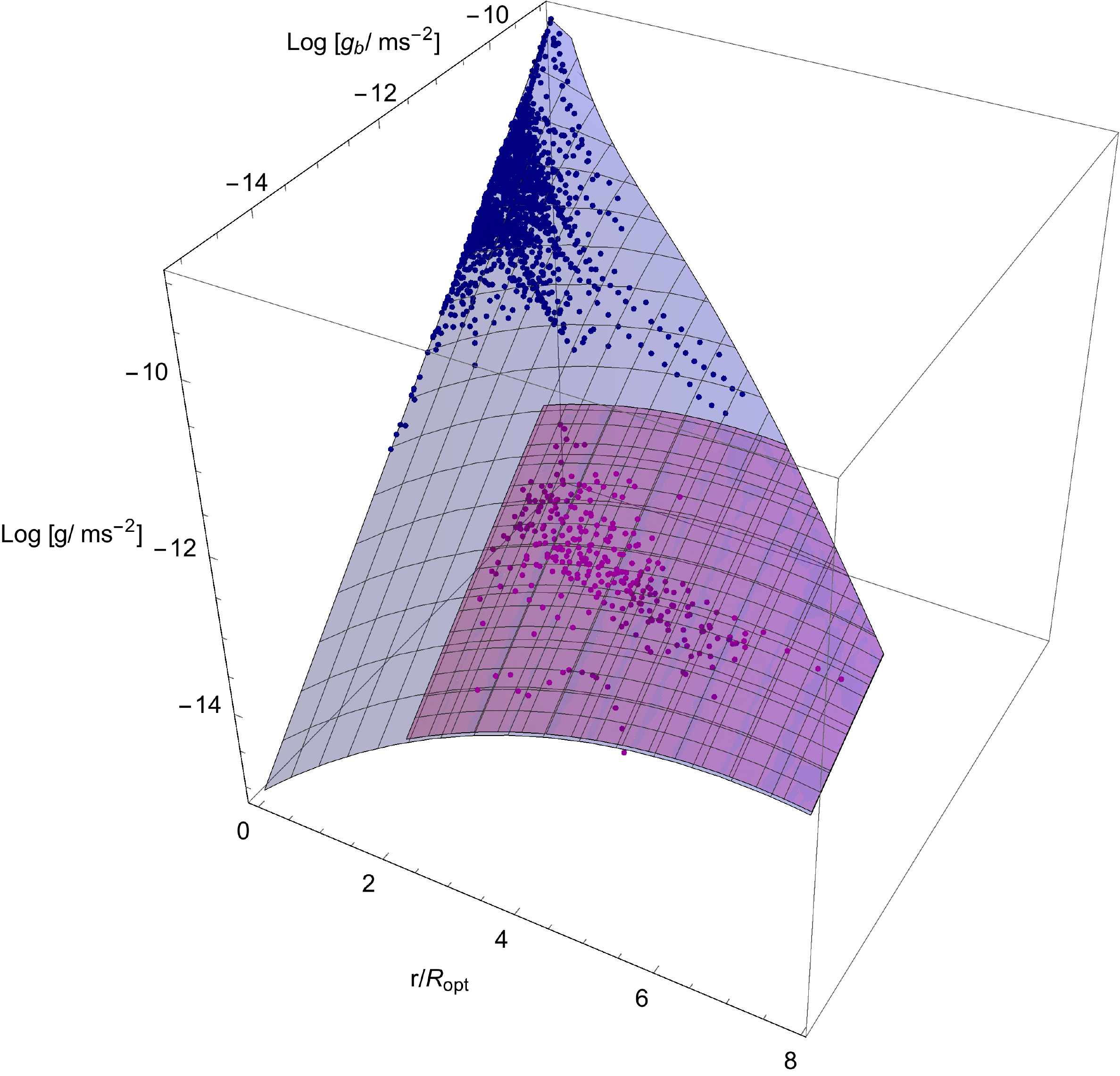}
\caption{The relationships among the total acceleration $g$, the baryonic acceleration $g_b$ and the normalised radii $r/R_{opt}$ for our two samples. The {\it magenta} and {\it blue points} refer to DD and LSB data, alongside with their best-fit {\it surfaces}.
 The LSBs measurements extend in the $Log \, g/ms^{-2}$ and $Log \, g_b/ms^{-2}$ range $\sim[ -12.5 \,, \,-9.0]$. The fitting surface well represent  also the very low accelerations data  discussed in the Appendix \ref{The extended $g - g_b$ plane}.}  
\label{3D_g_gb_x}
\end{center}
\end{figure}

The baryonic fraction $f_b$ as function of $r/R_{opt}$ for galaxies tagged  by $V_{opt}$ is given by:
\begin{eqnarray}
\label{Baryon_fraction}
\small
f_b(r/R_{opt}, V_{opt}) = \frac{V^2_{URC, \, b}(r/R_{opt}, V_{opt})}{V_{URC}^2(r/R_{opt}, V_{opt})}  \quad .
\end{eqnarray}  
See Fig. \ref{fb_plot_LSB}. Note that, going from 
the inner to the external radii and from the biggest to the smallest galaxies, the baryonic component becomes less and less relevant than the DM one. It is remarkable that  a very similar behaviour of  $f_b(r/R_{opt}, V_{opt})$  is found also in Spirals ~\citep{Salucci_2007, Lapi_2018}. 
\begin{figure*}[!t]
\begin{center} 
\includegraphics[width=0.43\textwidth,angle=0,clip=true]{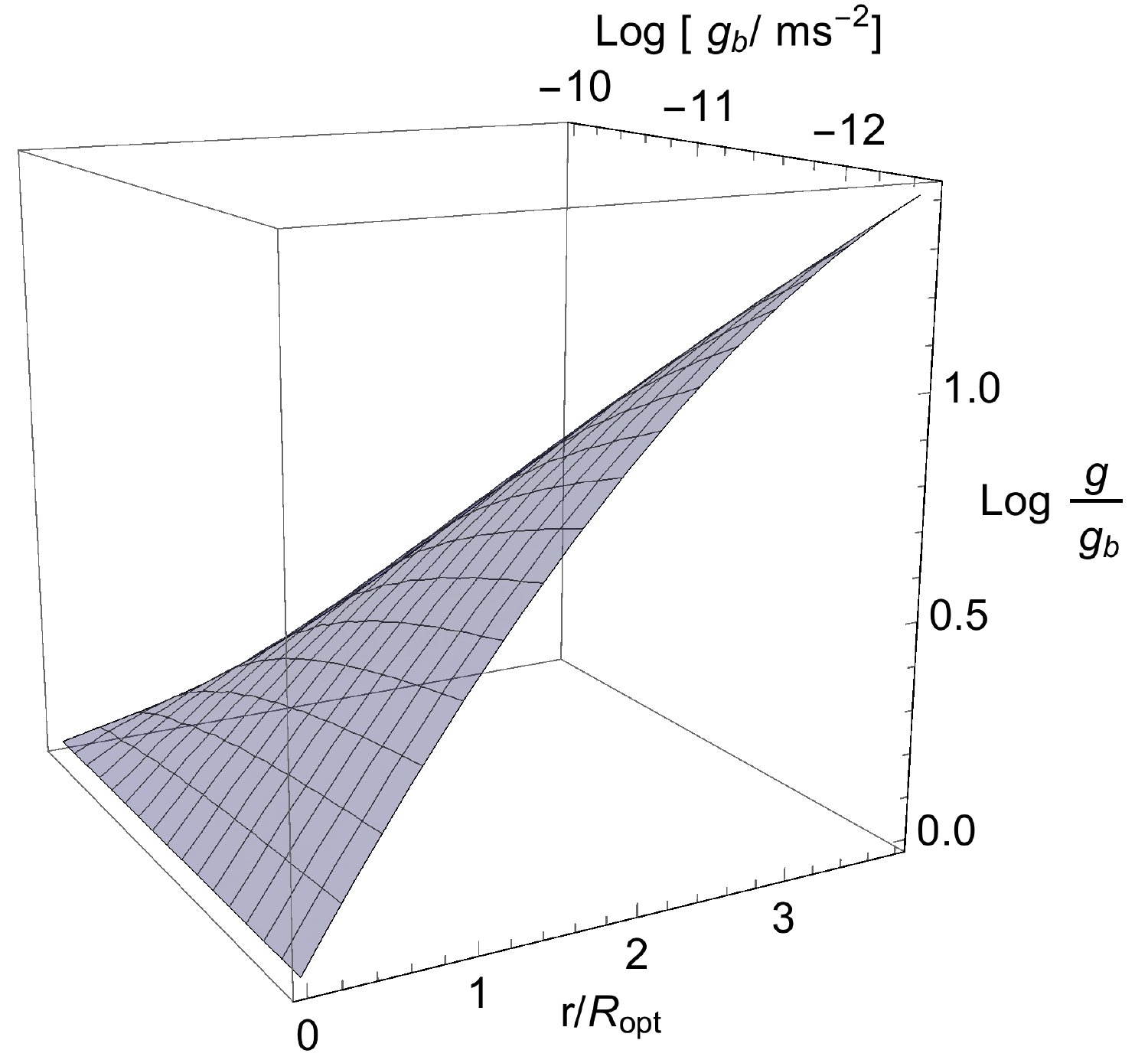}\includegraphics[width=0.63\textwidth,angle=0,clip=true]{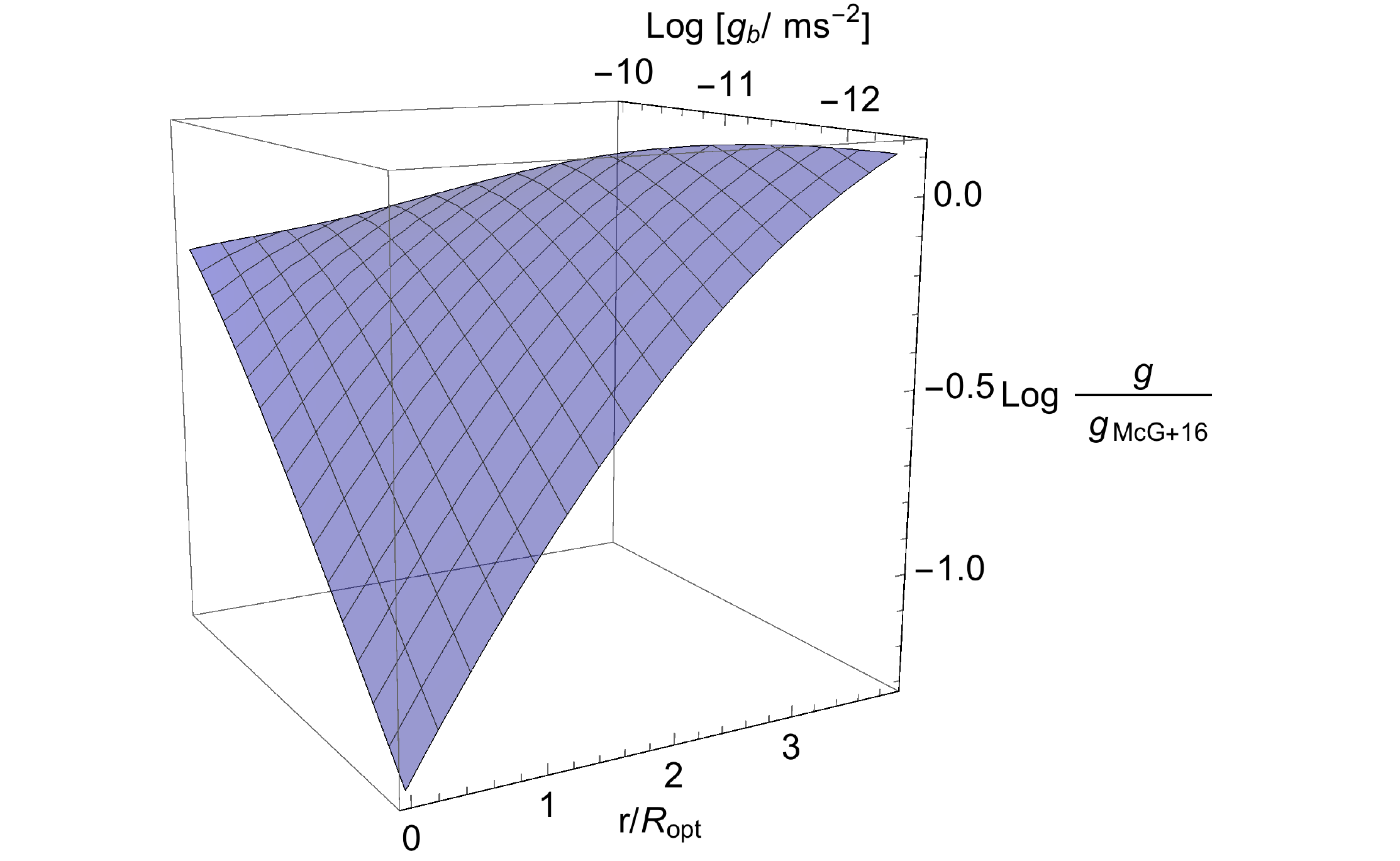}
\caption{The {\it surface in left panel} is given by the difference between the LSBs $GGBX$ relationship (Eq. \ref{Fitting_model_LSB}) and the Newtonian value $Log \, g_b$. The {\it surface in right panel} is given by the difference between the LSBs $GGBX$ relationship and the McG+16 relation (Eq. \ref{g_McG+16}).} 
\label{g_minus_gb}
\end{center}
\end{figure*}

Eq. \ref{Baryon_fraction}, recast in other terms, becomes: $V^2_{URC, \, b}(r)=f_b(r, V_{opt})V_{URC}^2(r)$ and, consequently, 
with Eq. \ref{gravity}, we have for each galaxy:
\begin{eqnarray}
\label{fb_relation}
g_b(r)=f_b(r, V_{opt}) \, g(r)    \quad .
 \end{eqnarray}   
Then, by summarising: in each galaxy with disc scale length  $R_{opt}/3.2$, rotation curve $V(r,R_{opt})$ with $V_{opt}$ tag  value, we have :
$g(r) = V^2(r)/r$ and
$g_b(r) = f_b(r)g(r)$, 
where $f_b(r)$ is the baryonic
fraction (hereafter, for semplicity, we drop the family tag $ V_{opt}$). 
Notice that $g(r)$ is totally observed, $g_b(r)$ has a part derived from the baryonic component to the rotation
curves obtained by the baryonic mass distribution.

\section{Results}\label{Test for $g$ and $g_b$}
\noindent
The emerging $g$ vs $g_b$ relationships, obtained for DD and LSB galaxies, are shown in Fig \ref{g_gb_dd_LSB_2D}.
We realise that the universality of the $g(g_b)$ relation, holding in normal spirals ~\citep{McGaugh_2016, Salucci_2018}
breaks down in our samples.  
The scatters of DD and LSB data with respect to the McG+16 relation are 0.17 dex and 0.31 dex respectively. 
This big discrepancy cannot be due to observational or systematical errors, in fact 
we have used high-quality rotation curves, so that the observational uncertainties on $V^2(r)$, leading to $g(r)$, are are less than
20\%. Systematical errors are present only on the quantities $g_b = f_b g $, due to $f_b$. From the modelling of the co-added rotation curves in Spirals, DD and LSBs, the quantity $f_b$ has
fitting uncertainties running from 10\% at higher luminosity to 30\% at lower luminosity. This implies that the uncertainties on $Log \,
g_b$ lay in the range between 0.13 dex and at most 0.19 dex. In this work, as those discussed in previous sections, the determination of $g$ and $g_b$ is not an issue. 
 It is important to note in Fig. \ref{g_gb_dd_LSB_2D} that there are many points strongly discrepant with respect to the McG+16 relation along {\it both} axes: in detail 
 1 dex on the $Log \, g_b$ axis and the same value on the $Log \, g$ axis, where our measurements can be considered almost
error-free. 

Let us stress that, as consequence of the method employed to derive $g_b$,  we cannot have: $g_b > g$; only when we consider the fitting uncertainties on $g_b$, we obtain that this quantity  can (sligthly) overcome $g$  in average by a  value of  $\sim 0.1$ dex at 2$\sigma$ level of uncertainty (see 
Appendix \ref{Small errors on $f_b$: small effect on the $g$ - $g_b$ plane}). This point is irrelevant for the scope of this paper.

The data relative to  the inner regions  of galaxies 
(red data) are the closest to the equality line $Log\, g = Log \,g_b$, while data relative to more external
regions
(blue data) of galaxies tend to depart from the equality line towards the region covered by McG+16 relation and then go beyond, with
$Log\, g > Log \,g_b$. 
This behaviour is intrinsically related to  the mass distribution in galaxies: 
the higher is the baryonic fraction $f_b$, the more $g$ is close to $g_b$, 
and reversely the lower is $f_b$, the more $g$ overcomes $g_b$.

\section{The universality of the $GGBX$ relationship}\label{The true universality of the $g$ vs $g_b$ relationship}
\noindent
It is evident that, in both DD and LSB samples, pairs of accelerations ($g$ ,\, $g_b$) residing at different radii  $r/R_{opt}$ don't overlap.
We realise that a relationship between $g$ and $g_b$ 
necessarily must involve also the position $x$, where the two accelerations are measured, and the Hubble type of the objects.
This is shown in our new 3D relationship, Eq. \ref{Fitting_model_LSB}, (hereafter $GGBX$ relation) among the  
$Log \,g - Log \, g_b - \, x $ quantities.
Starting from the McG+16 relation (in order to have a straightforward comparison), we added new terms to find 
 the best fitting model for LSB data. The best and simplest model that we found is:
\begin{eqnarray}
\label{Fitting_model_LSB}
 Log\, g_{_{LSB}} (x,\, Log\, g_b) = (1+a \, x) \, Log\, g_b    +  \,  b \; x \, Log \, [1-exp (-\sqrt{g_b(x)/\tilde{g}} ) ]  
+  \,c \, x \,+ \,d \, x^2    \quad ,
\end{eqnarray}   
where the fitting parameters $a, \, b , \, c, \,d $ assume the best-fit values -0.95, 1.79, -9.01, -0.05 respectively. The scatter of LSB data from the fitting surface is considerably reduced, down to 0.05 dex, i.e. to a sixth of the scatter from the McG+16 relation. 
Let us notice that the model used in Eq. \ref{Fitting_model_LSB} is just an empirical function used to fit the data that  
recovers $Log \, g \rightarrow Log\, g_b$ when $x \rightarrow 0$. Then the  number of free parameters of the $x$ part in the above relation expresses only our ignorance of the actual functional form of the relationship and  not the fact that  the $g (g_b,x)$ surface is not smooth and of negligible  thickness.  

In the case of DD galaxies, by simply applying translations and/or dilatations to Eq. \ref{Fitting_model_LSB} along the three involved axes, we obtain the following best fitting model:
\begin{eqnarray}
\label{Fitting_model_DD}
Log\, g_{_{DD}} (x,\, Log\, g_b) =   
 Log\, g_{_{LSB}}\left(\frac{x}{l}  + \, h,\, \frac{Log\, g_b }{ m } + \, n \right) + \, q   \quad .
\end{eqnarray}   
We found a perfect fit of the data when the fitting parameters $ l, \, h , \,  m , \, n, \, q  $ assume the best-fit values 0.49,   2.41 , 
  0.74  ,   1.72  ,  1.19  respectively. 
The scatter of DD with respect to the fitting surface is considerably reduced, with a value of 0.03 dex, i.e. about a fifth of the scatter from the McG+16 relation. 
\\
\\
We show in Fig. \ref{3D_g_gb_x_a} the DD and LSB data in the $g-g_b-x$ space, with their best fitting surfaces from Eq. \ref{Fitting_model_LSB}-\ref{Fitting_model_DD}.
The result is extremely remarkable. It shows a precise relation linking the total and baryonic acceleration, the galactocentric distance $x \equiv r/R_{opt}$ and even the morphology of galaxies.
The scatter of both LSB plus DD data (after the translation and dilations given by the parameters $ l, \, h , \,  m , \, n, \, q  $; see Fig. \ref{3D_g_gb_x}) from the $GGBX$ surface is only  0.05 dex, about a sixth of their scatter from the McG+16 relation (0.29 dex); moreover it is also lower then the scatter of 0.13 dex of McG+16 sample from McG+16 relation.  
The statistical significance is overwhelming, but its physical meaning is not immediate.
Let us stress that  the data $g,\, g_b, \,x$ form, for LSBs and DDs, two very thin {\it surfaces}
that can be overlapped through a simple coordinate transformation. Again,  the number of the fitting parameters reflects 
our ignorance of the analytical representation of the $g (g_b,x)$ relation, not the statistical relevance of the surfaces defined by data.  

\subsection{Understanding the $GGBX$ relationship}\label{Understanding the $GGBX$ relationship}
\noindent
Our relationship deviates both from the Newtonian and from the McG+16 relationship. In particular, by considering the LSBs, i.e. our most numerous sample, we observe that:

$i)$ the deviation from the Newtonian relation is more evident at larger galactocentric radii and for smaller $g_b$ values. 
See the left panel of Fig. \ref{g_minus_gb}, which shows the difference $Log\, g_{_{LSB}} (x,\, Log\, g_b) - Log\, g_b$;
 
$ii)$ the deviation from the McG+16 relation is particularly evident at smaller galactocentric radii and for smaller $g_b$ values.
See the right panel of Fig. \ref{g_minus_gb}, which shows the difference $Log\, g_{_{LSB}} (x,\, Log\, g_b) - Log\, g_{_{McG+16}}$.

We highlight that these results are related to the mass distribution in galaxies:  any $g_b(r)$ corresponds to very different values of $g(r)$ according to the tag velocity $V_{opt}$ (or luminosity), the normalised radius $r/R_{opt}$ and the Hubble Type of the galaxy in question. This is consequence of the fact that $g_b(r) = f_b(r) g(r)$ and that $f_b(r)$, related to the mass distribution in galaxies, depends on the tag velocity $V_{opt}$ (or luminosity), the normalised radius $r/R_{opt}$ and the Hubble Type of the galaxy in question (Fig. \ref{fb_plot_LSB}).

It is worth to show how all the above results, including the disagreements with  McG+16, are evident when we plot the GGBX relation in  {\it individual }objects (see Appendix \ref{Single galaxies analysis}).

In conclusion, straightforward facts are that:

(i) the same values of the pairs $(g, \, g_b)$ found in the outer region of big spirals are replicated in the inner region of small spirals, provided that approximately $r \geq R_d$.
This explains the genesis of McG+16 and S18 findings;

(ii) given one spiral and one LSB, both with the same $V_{opt}$ and then very similar $f_b(x)$, they can show very different $f_b(r)$ in physical radial units. This happens because LSBs usually have much more extended $R_D$ than spirals (see Fig. 9 in ~\citep{DiPaolo_2018}).
Thus, $f_{b,_{LSB}}(r) > f_{b,_{spiral}}(r)$. 
Then, at fixed value of $g_b$,  very different values  of $g$ can correspond, and vice-versa.
This mainly explains the failure of the McG+16 relation in LSBs.

\section{Conclusion}\label{Conclusion}
\noindent
The two accelerations relationship (eq. \ref{g_McG+16}) by McG+16 has attracted a large interest. It is claimed and thought that it provides crucial evidence about the issue of dark matter. In this work, we have investigated the $g_b - g$ relationship (found by McG+16 for a sample dominated by normal spirals), in the recent sample of 36 Dwarf Discs and 72 LSB galaxies, whose optical velocities span from 
$\sim 17 \, km/s$ to $\sim 300 \, km/s$, covering the full population of galaxies sizes and luminosities.  We analyzed overall 1904 velocity data and modeled them by involving an exponential stellar disc, a Burkert dark matter halo density profile ~\citep{Karukes_2017, deBlok_2002} and, in particular, we also considered the Tonini et al HI discs ~\citep{Tonini_2006} in DD galaxies and a bulge component in larger LSB galaxies ~\citep{Karukes_2017, DiPaolo_2018}. 
Then, we have derived the 1904 ($g_b$, $g$) pairs in the same way of McG+16 with the difference that the disc masses are obtained kinematically. 
This difference of methods, however leads to estimates of the disc masses that agree within their uncertainties. The great 
discrepancy between the McG+16 relationship and ours does not arise from the adopted values of the stellar disc + HI disc masses. 

In our objects $Log \, g/ ms^{-2}$ lays in the range between -14.5 and -9. On the other hand, the unbinned data $Log \, g/ ms^2$ in the McG+16 relationship range between -11.4 and -8.
The results of our tests involving the DD and LSBs samples show empirically that the radial acceleration $g$ in galaxies is not simply a universal function dependent on the baryonic acceleration  $g_b$ (as claimed by McG+16 in eq. \ref{g_McG+16}), but also depends on the galactic radius expressed  in normalised units $r/R_{opt}$. 

The emerging relationship  mirrors the properties of the  DM in galaxies, whose fraction changes along the galactic radius, becoming more dominant  on the baryonic one in the external regions, in a way which depends on the morphology and the luminosity of the galaxy (Fig. \ref{fb_plot_LSB}) ~\citep{Persic_1996}.

 For each sample, we have established a universal relation $ g = f(g_b, x)$ (that we call $GGBX$ relationship), with $x$ the normalised radius with respect the optical radius $R_{opt}$.
Moreover, we can go from DD relationship to the LSB one by means of translations and/or dilatations of the three involved variables. The individual average scatter around  these $GGBX$ new surfaces (created  by $g$, $g_b$ $x$ data) is remarkably reduced with respect to that around  to the McG+16 relation, more precisely it becomes a fifth and a sixth for DD and LSB galaxies data, respectively.  

Our relationship deviates both from the Newtonian and from the McG+16  2D relationship. In particular, 
the deviation from the Newtonian relation is more evident at larger galactocentric radii and for smaller $g_b$ values, while 
the deviation from the McG+16 relation is particularly evident at smaller galactocentric radii and for smaller $g_b$ values.

It is worth saying that the results are intrinsically related to the {\it mass distribution in galaxies}, i.e. to the variation of the baryonic 
fraction $f_b$ along the galactocentric radius and on the fact that it changes when we consider galaxies of different luminosity and different Hubble Type. This implies that, when considering different galaxies, a same value of $g_b$ can be found at very different radii $r$ and can correspond to very different values of $g$. This is the main explanation of the discrepancy among LSBs, DD and Spiral galaxies considered in McG+16 and S18.  

In this paper a new relation among the dynamical quantities in disc galaxies has emerged. The further investigation of the origin of such relation 
and the consequences in single objects will be shown in another next paper in preparation by Di Paolo et al. (2018).

In conclusion, we find that the $GGBX$ relationship (Eq. \ref{Fitting_model_LSB}-\ref{Fitting_model_DD}) is universal 
and  framed in the DM + Newtonian gravity scenario. We point out that this relation  stems out of  the properties of   $V^2(x)$ and 
$f_b(x)$ . Therefore,  it does not pose issues to the $\Lambda CDM$ + baryonic 
feedback scenario. 

Crucial properties of the DM are instead unlikely to come from the $g-g_b$ relationship, in fact the DM halo density profile is $\rho(r)= 1/(4 \pi G r^2)  \;  \frac{d}{dr} [ g(r) r^2 (1- f_b(r))]$ and crucially depend on quantities not present in the $g-g_b$ relationship: e.g.
$dV(r)/dr$, $V^2(r) \,df_b(r)/dr$. 
Whether our $GGBX$ compacts all the structural properties of DM halos will be left to a further work (Di Paolo et al. (2018) in prep.).

\section*{acknowledgments}

\noindent 
We thank F. Nesti, A. Lapi, L. Danese and A. Erkurt for useful discussions. We also thank G. Costa for helps that have improved the presentation of the results of this paper.
%
%

\appendix

\section{The extended $g - g_b$ plane}\label{The extended $g - g_b$ plane}
\begin{figure*}[!h]
\begin{center} 
\includegraphics[width=0.7\textwidth,angle=0,clip=true]{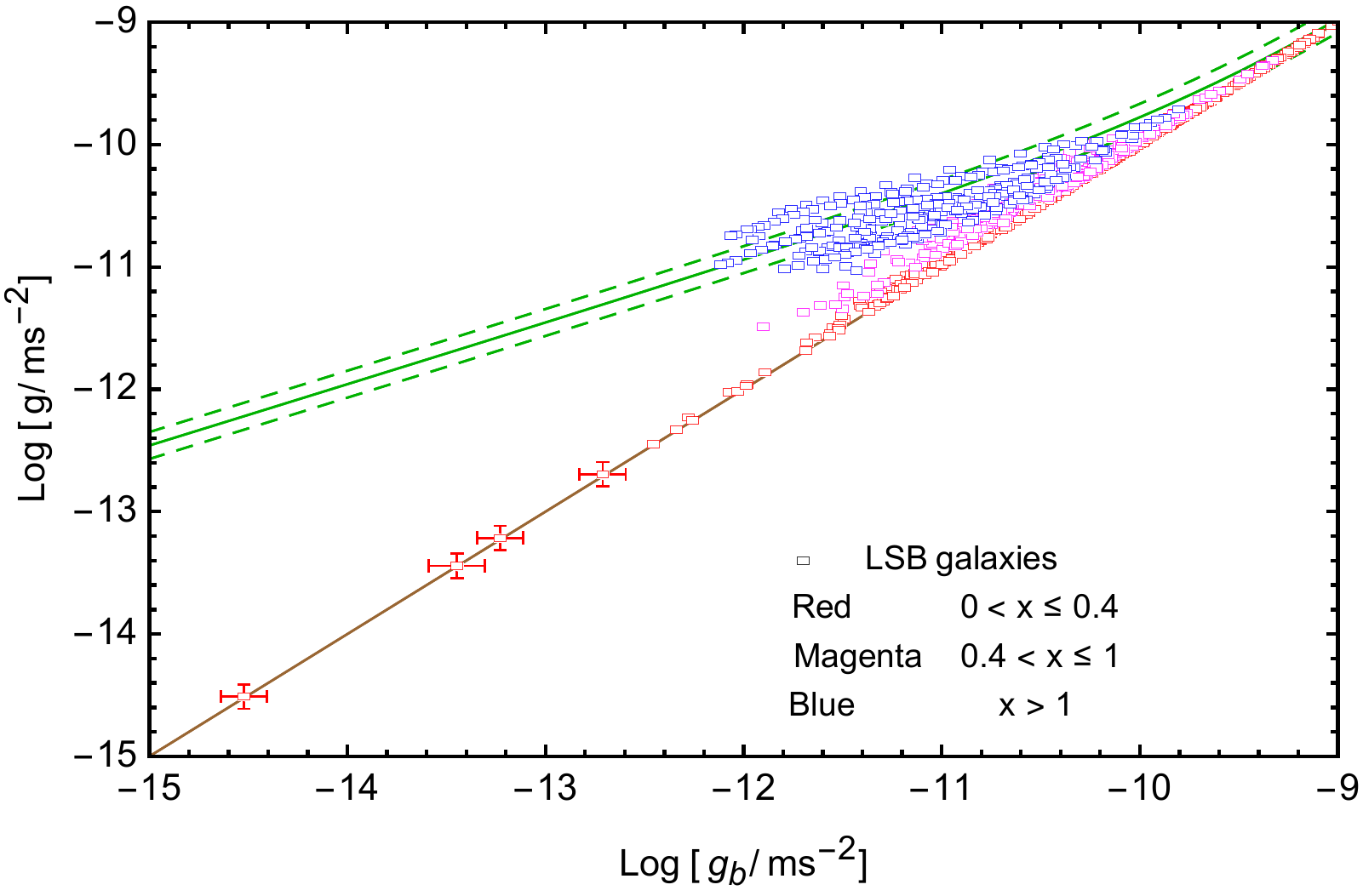}
\caption{Relationship between the total acceleration $g$ and its baryonic component $g_b$, for LSB data. $x=r/R_{opt}$. 
The figure is analogous to Fig. \ref{g_gb_dd_LSB_2D}, but also includes data till the lowest values of $Log \, g$ and $Log \, g_b $.
The 4 "special points" with very small values of $Log \, g$ - $Log \, g_b$ are shown with their $1   \sigma$ uncertainties. 
}  
\label{g_gb_extension}
\end{center}
\end{figure*}

\noindent 
 For completeness, we show all the LSBs data in Fig. \ref{g_gb_extension}, in order to highlight the extension of 
$Log \, g$ and $Log \, g_b $ values to $\sim -14.5$ (with the argument expressed in $m/s^2$).
We highlight that, originally, we had 1605 data for the LSB galaxies.  4 "special points" of them have very  low values of $Log\,g$ and $Log \, g_b$  laying in the range [-14.5, -12.5]. See Fig. \ref{g_gb_extension}.
These data strongly  support our result shown above, i.e. 
the discrepancy of LSB accelerations  from the McG+16 relationship, however, we keep them separately from the rest of the data because
 they are too few to cover their wide magnitude range (only 4 points in a range of 2 dex).

\section{The Universal Rotation Curve (URC) method}\label{TheURC method}
\begin{figure*}[!h]
\begin{center} 
\includegraphics[width=0.33\textwidth,angle=0,clip=true]{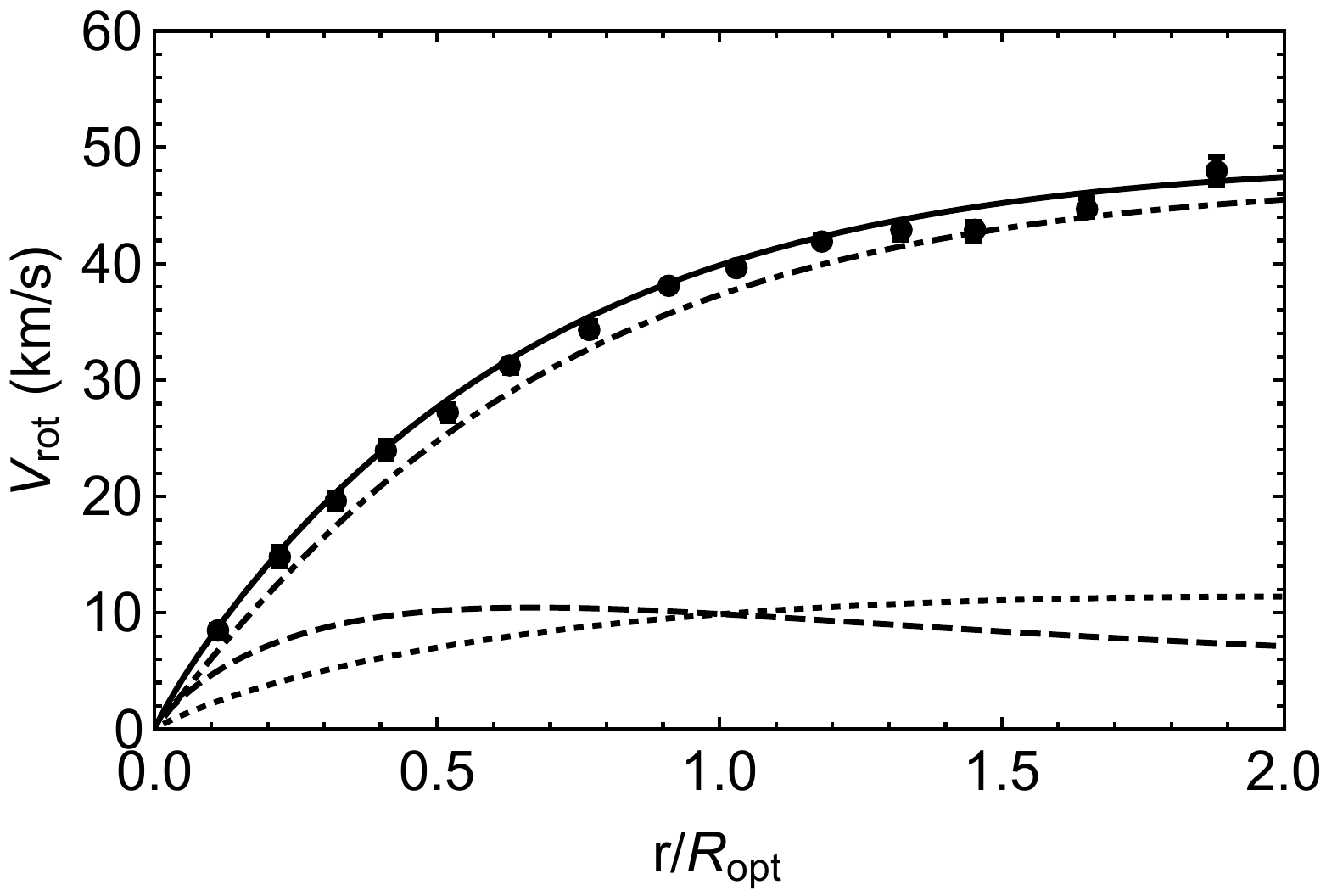}\includegraphics[width=0.323\textwidth,angle=0,clip=true]{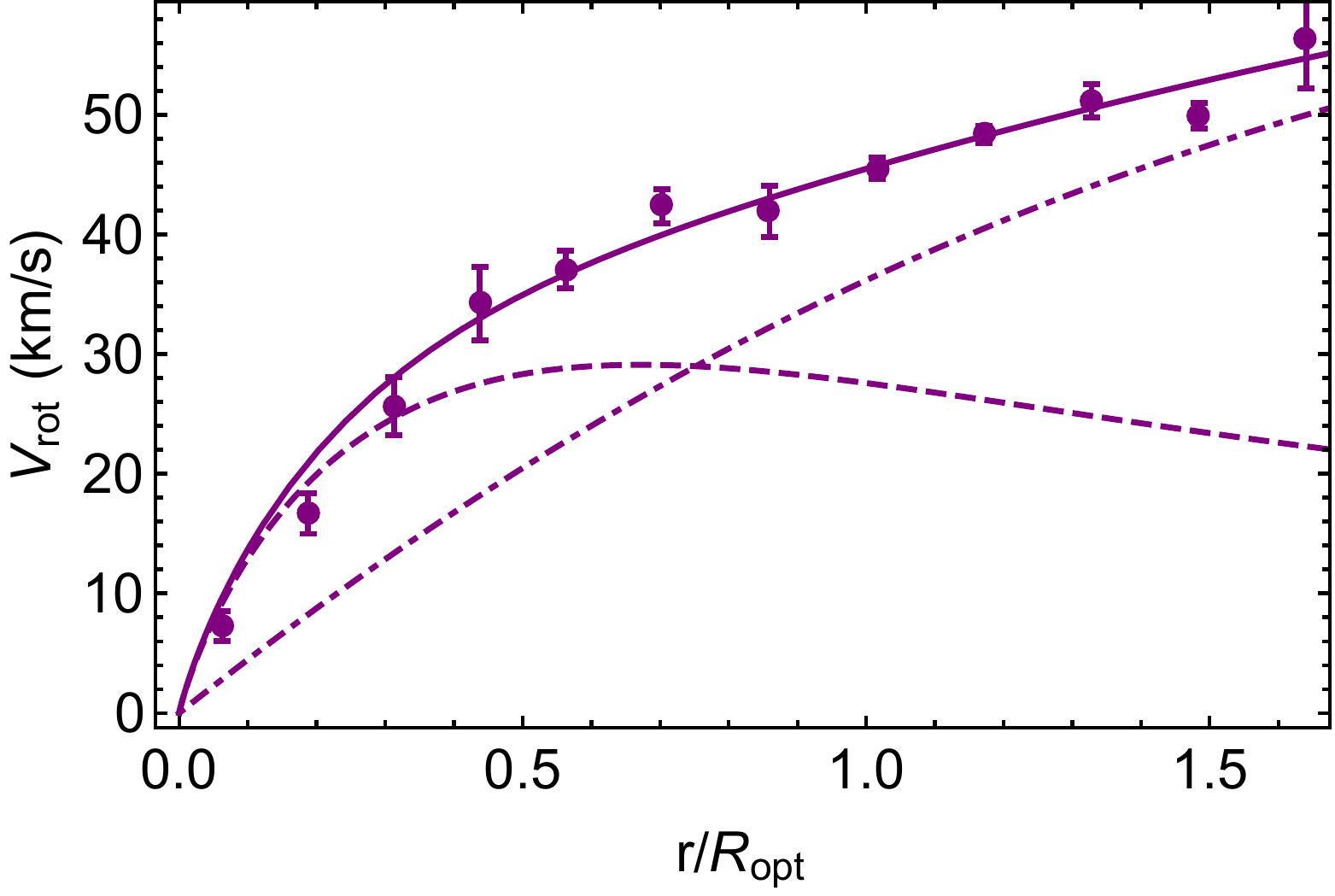}\includegraphics[width=0.33\textwidth,angle=0,clip=true]{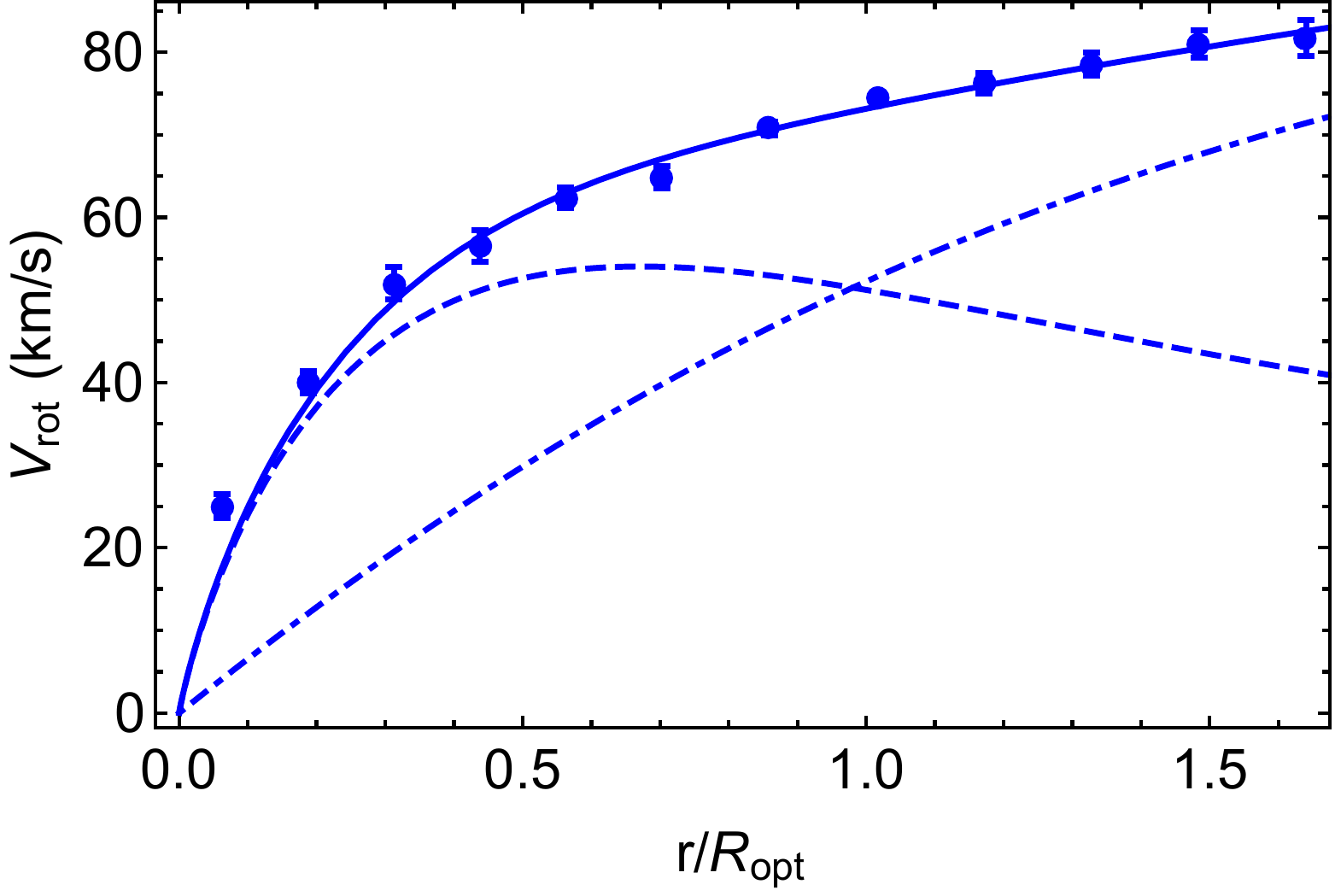} \\
\includegraphics[width=0.33\textwidth,angle=0,clip=true]{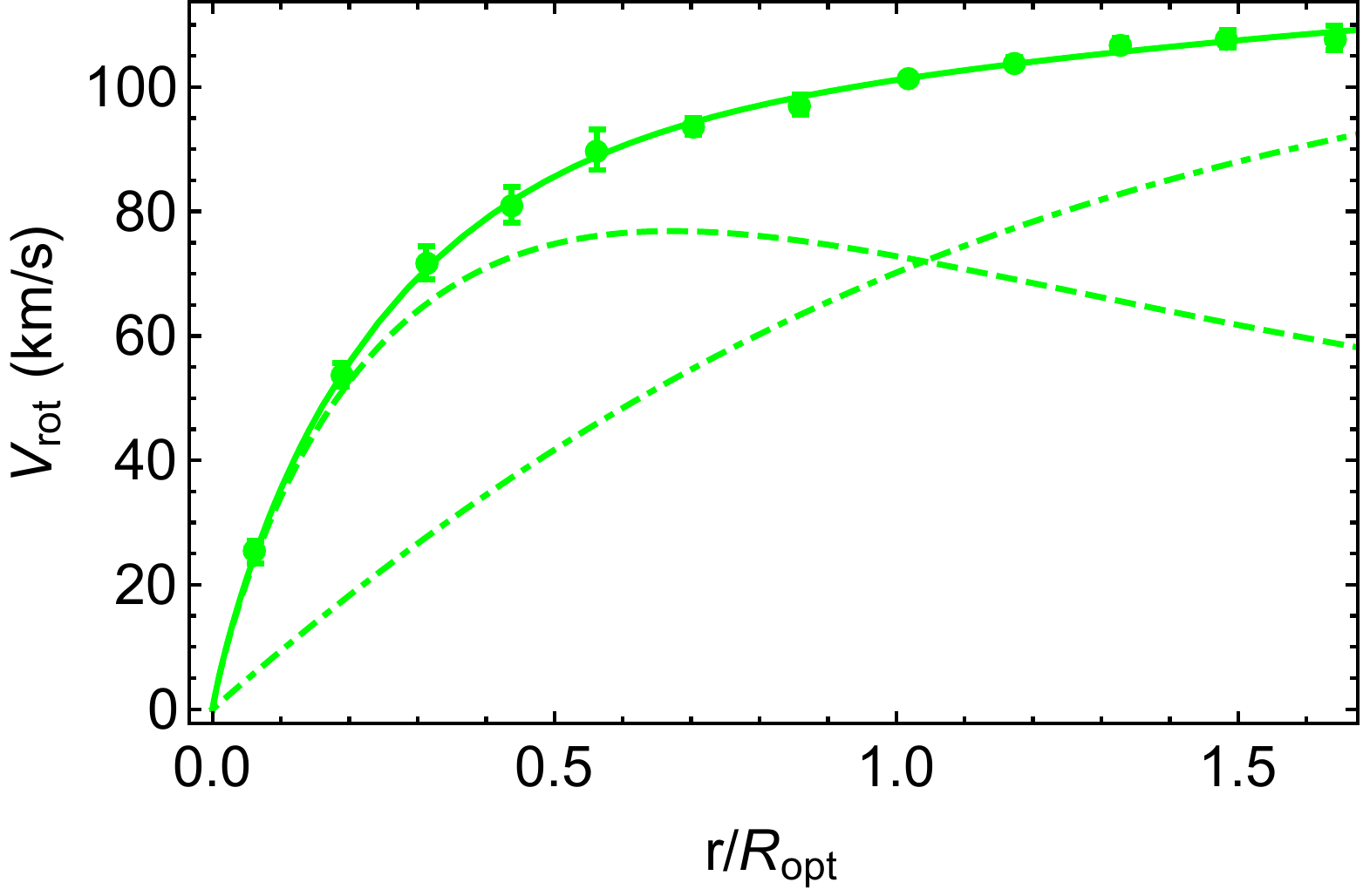}\includegraphics[width=0.323\textwidth,angle=0,clip=true]{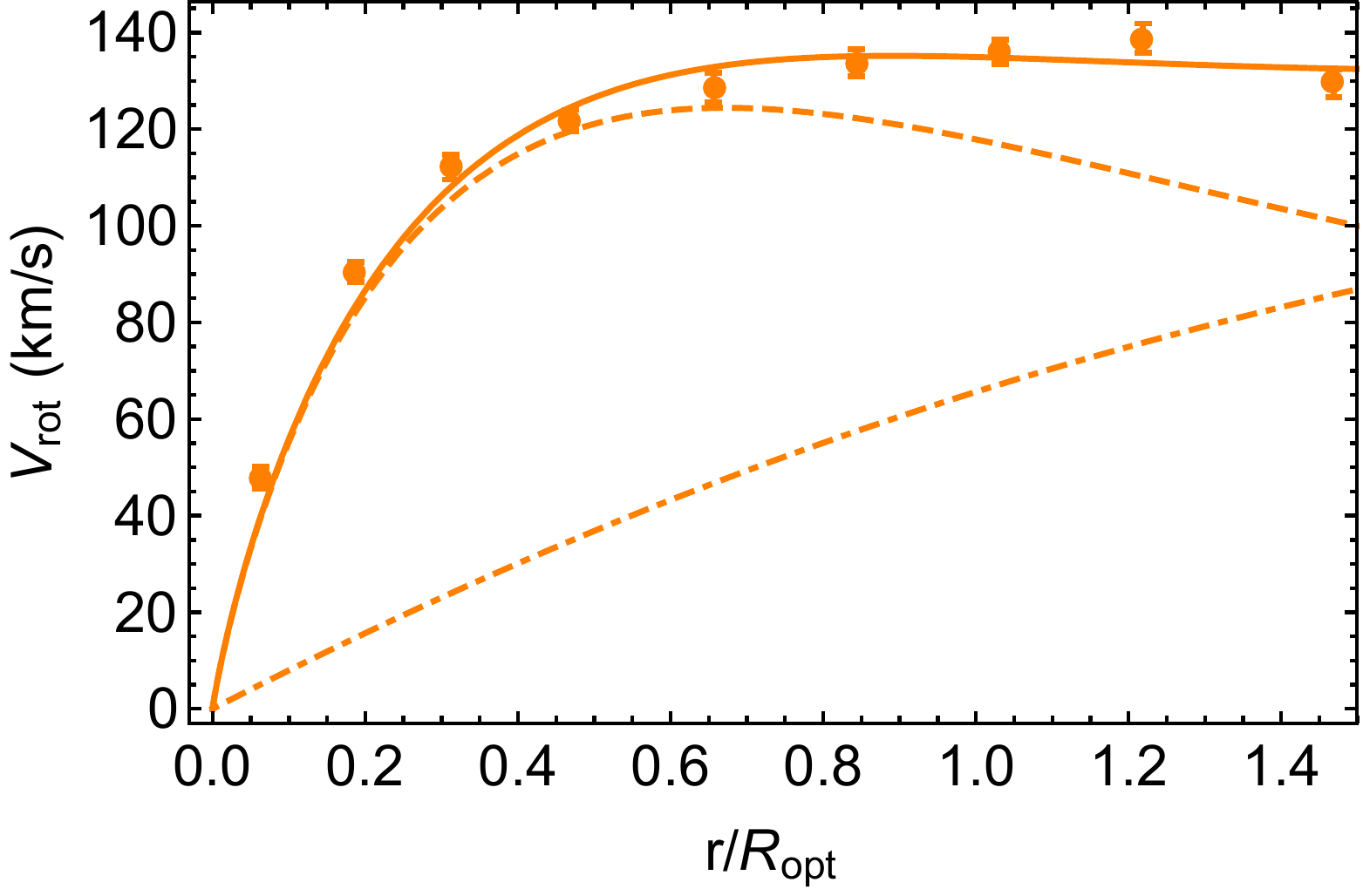}\includegraphics[width=0.329\textwidth,angle=0,clip=true]{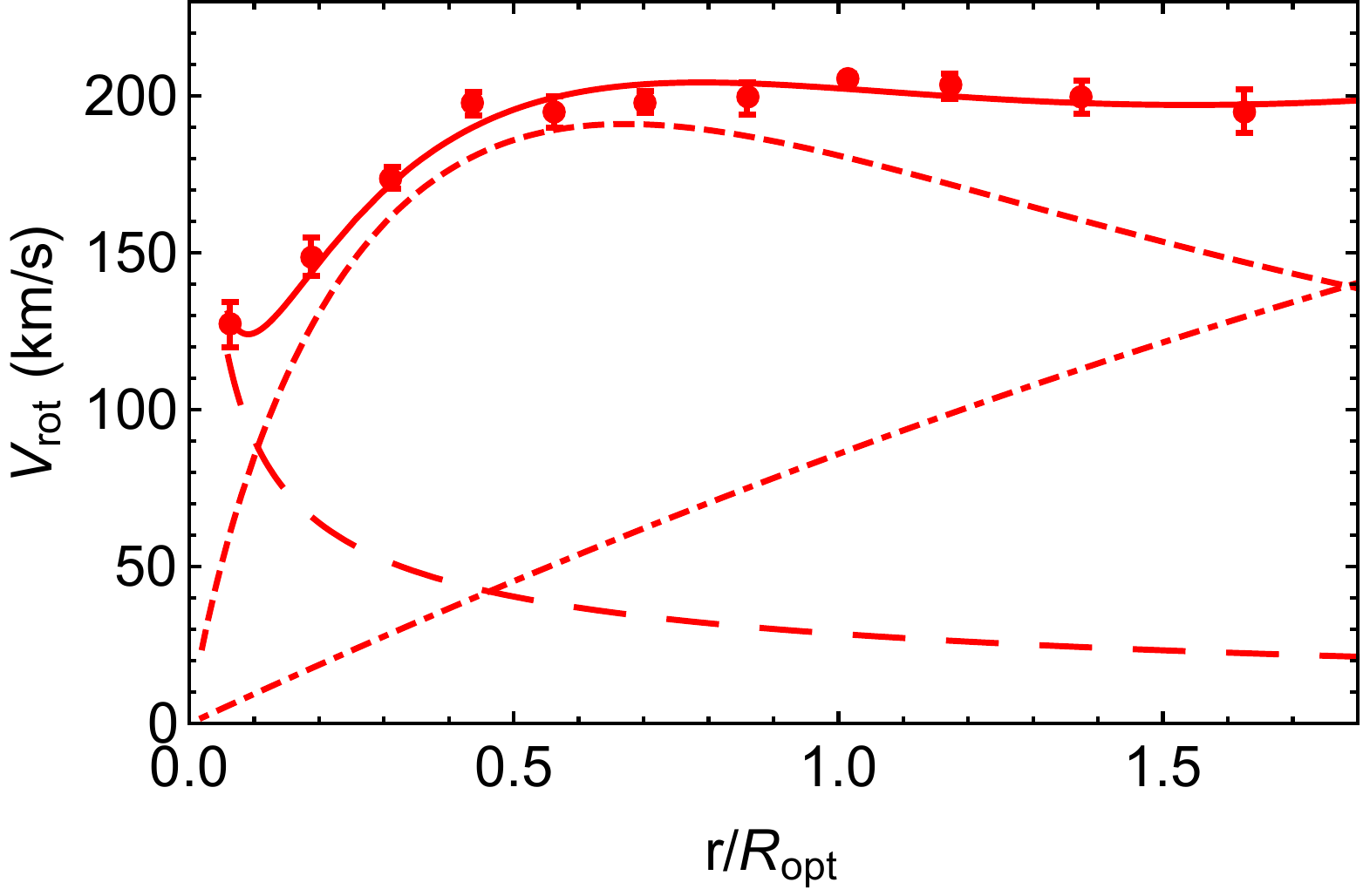}
\caption{Best fit URC velocity models of the co-added RCs of the unique velocity bin representative of DD galaxies ({\it black}, with $\langle V_{opt} \rangle  = 40 \, km/s$) and of the five  velocity bins, representative of the LSBs ({\it purple, blue, green, orange} and {\it red} with: $ \langle V_{opt} \rangle = 43, \,73, \, 101, \, 141, \, 206 \, km/s$). The {\it dashed, dotted, dot-dashed, long-dashed} and {\it solid lines} are the stellar disc, HI disc, dark matter, bulge and total contributions to the circular velocities, respectively.}  
\label{Velocity_Bin}
\end{center}
\end{figure*}
\noindent
The URC is derived, firstly,   by   luminosity/optical velocity  and  normalized radial binning of a large number of  individual rotation curves that yield suitable
co-added rotation curves $V_{co-add}(x \footnote{$x=r/R_{opt}$} ,\lambda \footnote{$\lambda$ is equal to $M_k$ or $V_{opt}$, (i.e.
$\lambda$ is the galaxies family identifier).})$, see for details ~\citep{Persic_1996, Salucci_2007}. For the present work:
 the DD galaxies are grouped in a single family ~\citep{Karukes_2017}, the LSB
galaxies are grouped in five families, each with increasing tag average  velocity $\langle V_{opt}\rangle$. 

The co-added curves (RCs) are very well reproduced by a suitable analytical velocity model  that we call $V_{URC} (x,\lambda)$ (see
~\citep{Karukes_2017, DiPaolo_2018}). \footnote{ For our objects  we know  the values of their $R_d$, so that  we can express the URCs in term of their physical radial units $V_{URC}(r)$.}

The URC method has been  applied,  so far,  to Spirals, LSB and dwarf discs. It   consists in the sum in quadrature of four terms, $V_{d,
URC}$ , $V_{HI, URC}$, $V_{bu, URC}$, $V_{h, URC}$, each of them  describing the contribution from the stellar disc, the HI gaseous disc,
the central bulge and the dark halo.  Then: 
\begin{eqnarray}
\label{V_Model}
V_{co-added}^2(x,\lambda)  \, \simeq  V_{URC}^2(x,\lambda) = V_{d, URC}^2(x,\lambda) + 
                               V_{HI, URC}^2(x,\lambda)  +V_{bu, URC}^2(x,\lambda)+V_ {h, URC}^2 (x,\lambda)  \quad ,
\end{eqnarray}   
\noindent
where the l.h.s. are the co-added RCs and the r.h.s is the analytical model with which we fit the former.

For simplicity, hereafter we drop the tag "URC" in the model velocity components.

\noindent
$V_{URC}(x,\lambda)$
 fits extremely well all $V_{co-added}(x,\lambda)$ (see Fig. \ref{Velocity_Bin}) of spirals ~\citep{Persic_1991, Salucci_2007, Persic_1996}, DD ~\citep{Karukes_2017} and LSB ~\citep{DiPaolo_2018}, and provides us with 
 an accurate analytical  representation of the individual rotation curves.

The stellar component is described by means of  the well-known exponential disc ~\citep{Freeman_1970} with 
surface density profile 
$\Sigma _D (r) = \frac{M_d}{2\pi R_d^2}\,  exp(-r/R_d) $. 

Caveat the distance of the galaxy, the gas contribution is known from observations ( e.g. see ~\citep{Evoli_2011}).
This component is described as it follows: the total mass is obtained from the  21-cm  flux and its radial distribution is given by 
$\Sigma _{HI} (r) =  \frac{M_{HI}}{2\pi (3 \, R_d)^2}\,  exp(-r/3 \, R_d) $ ~\citep{Tonini_2006, Evoli_2011, Wang_2014}. 
Then:
\begin{eqnarray}
\label{V_stellar}
V_d^2(r) = \frac{1}{2} \frac{G\, M_d}{R_d} (3.2 \, r/R_{opt})^2 (I_0 K_0 - I_1 K_1)   \;\;  ; \; \;\;\;\;\;\;\;  
V^2_{HI}(r)=\frac{1}{2}\frac{G M_{HI}}{ 3 R_{D}} ( 1.1 \,r/R_{opt})^2 (I_{0} K_{0}-I_{1} K_{1})     
\end{eqnarray}   
where $M_d$ is the stellar disc mass, $M_{HI}$ is the gaseous disc mass (correcting by a factor 1.3 in order to account for the He 
abundance),  $I_n$ and $K_n$ are the modified Bessel functions computed at $1.6 \,x$  and $0.53 \, x$ for the stellar and the gaseous disc respectively.

Let us notice that, in LSBs,  
 the gas contribution to the circular velocity is  negligible for the scope of this paper (see also Appendix \ref{The gas effect on the $g$-$g_b$ plane}). 

\noindent
 In the largest velocity  bin of LSBs, 
in the URC model  we have included a bulge component by adopting:
\begin{eqnarray}
\label{Bulge_velocity}
V^2_{bu}(r)= \alpha _b V^2_{in} (r/r_{in})^{-1}  \quad ,
\end{eqnarray}   
where $V_{in}$ and $r_{in}$ are values referred to the innermost circular velocity measurements and $\alpha _{b}$ is a parameter varying from $0.2$ to $1$ (see e.g. ~\citep{Yegorova_2007}). 

Therefore, for bulgeless DD galaxies we assume, as baryonic contribution, $V_b^2 (r) = V_d^2(r) + V_{HI}^2(r)$, while for the LSBs we assume $V_b^2 (r) = V_d^2(r) $ for the four galaxies families (velocity bins) characterised by the smallest $V_{opt}$ and $V_b^2 (r) = V_d^2(r) + V_{bu}^2(r)$
for galaxies with the largest $V_{opt}$ ~\citep{Salucci_2000, Das_2013}.
\\
\\
For the DM halo velocity contribution we adopt the cored Burkert profile ~\citep{Burkert_1995}: 
\begin{eqnarray}
\label{Burkert}
 V^2_{h}(r)   =  2 \pi G \rho_{0} \frac{R_{0}^3}{r}  [\ln(1+r/r_0)   
                     -tg^{-1}(r/r_0)+0.5\ln (1+ (r/r_0)^2 ) ]    
\end{eqnarray}   
where $\rho _0$ is the central mass density and $r_0$ is the core radius.
\\
\\
By resuming, 
the co-added rotation curves $V_{co-added}$ are very well fitted by $V_{URC}$ (see Fig. \ref{Velocity_Bin}) and the best fitting parameters $M_d, \,  \alpha _b, \,  \rho_0, r_0$ result all as a function of $\lambda$ ($V_{opt}$ or $M_K$). We direct the interested reader to 
~\citep{Karukes_2017, DiPaolo_2018}.

\section{The HI component effect on the $g$-$g_b$ plane}\label{The gas effect on the $g$-$g_b$ plane}
\begin{figure}[!h]
\begin{center} 
\includegraphics[width=0.33\textwidth,angle=0,clip=true]{Velocity_1.pdf}
\includegraphics[width=0.33\textwidth,angle=0,clip=true]{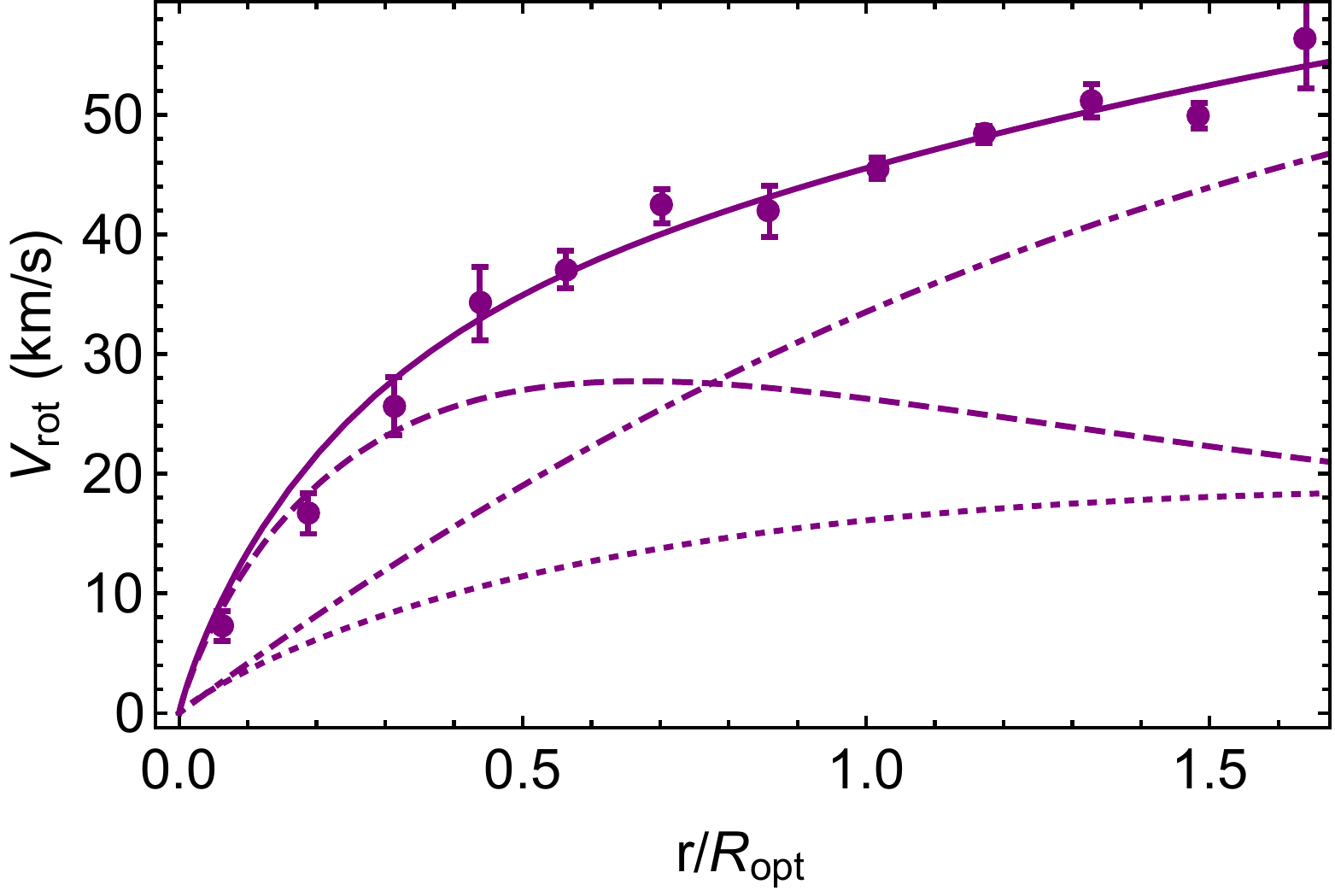}\\
\includegraphics[width=0.45\textwidth,angle=0,clip=true]{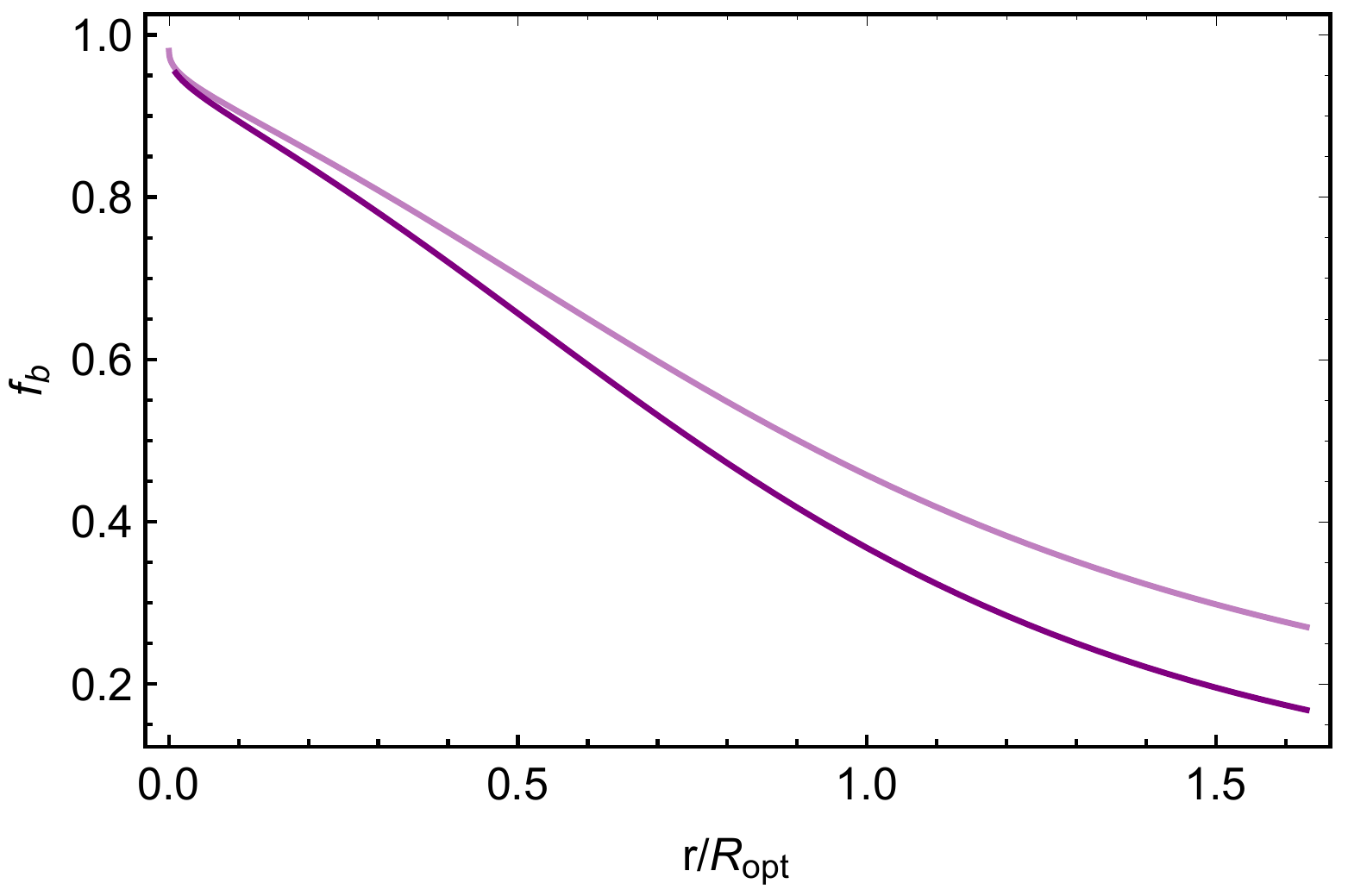}
\end{center} 
\caption{{\it Upper panels}: I velocity bin ({\it family}) rotation curve fitted without and with gas. The {\it dashed, dot-dashed, dotted} and {\it solid lines} stand for the stellar disc, the DM halo, the gaseous disc and the total contributions to the rotation curve, respectively.
{\it Bottom panel}: baryonic fraction without gas ({\it dark purple}) and with gas ({\it light purple}).}
\label{I_velocity_bin}
\end{figure}
\begin{figure}[!t]
\begin{center} 
\includegraphics[width=0.6\textwidth,angle=0,clip=true]{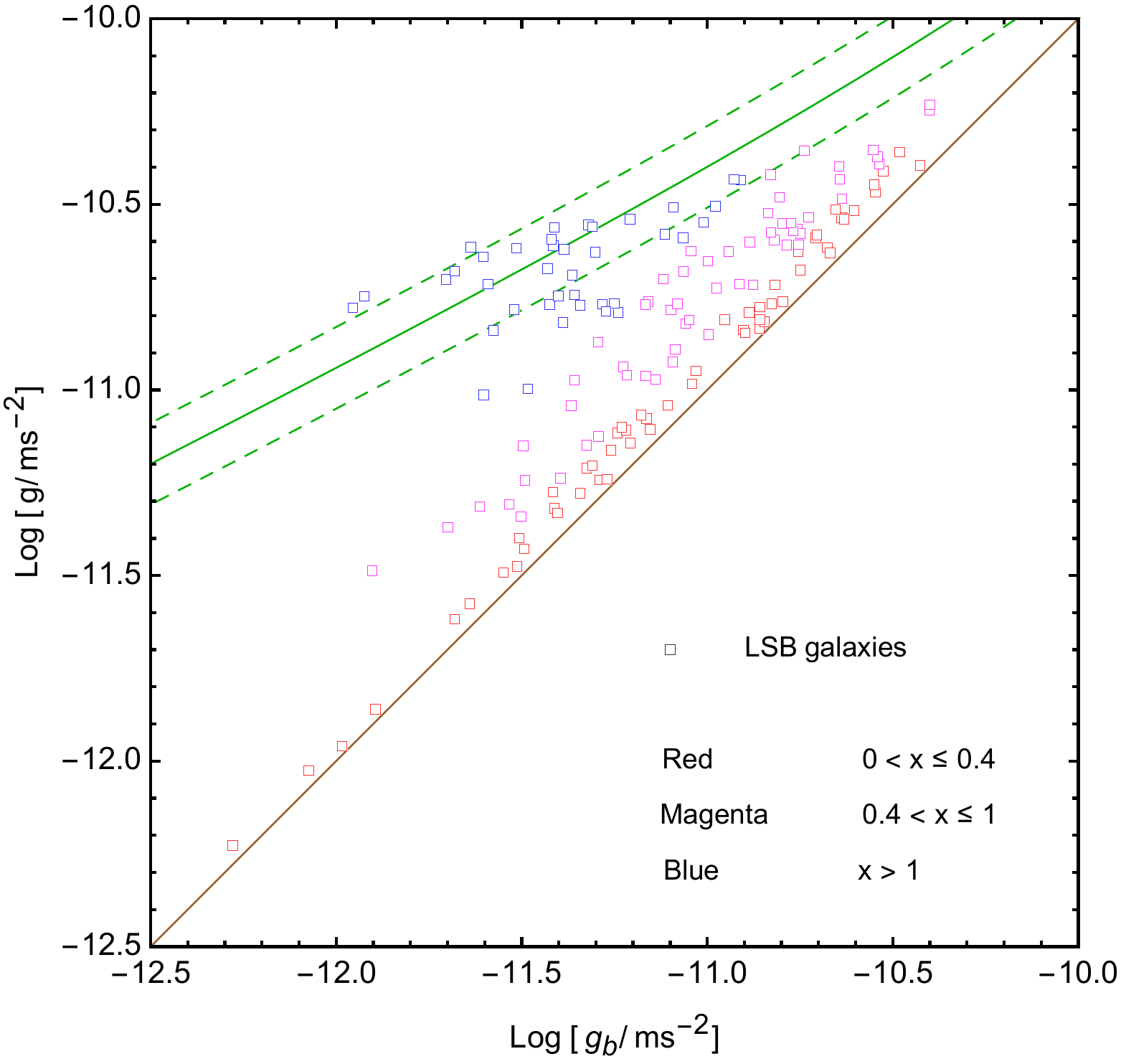} \\
\includegraphics[width=0.6\textwidth,angle=0,clip=true]{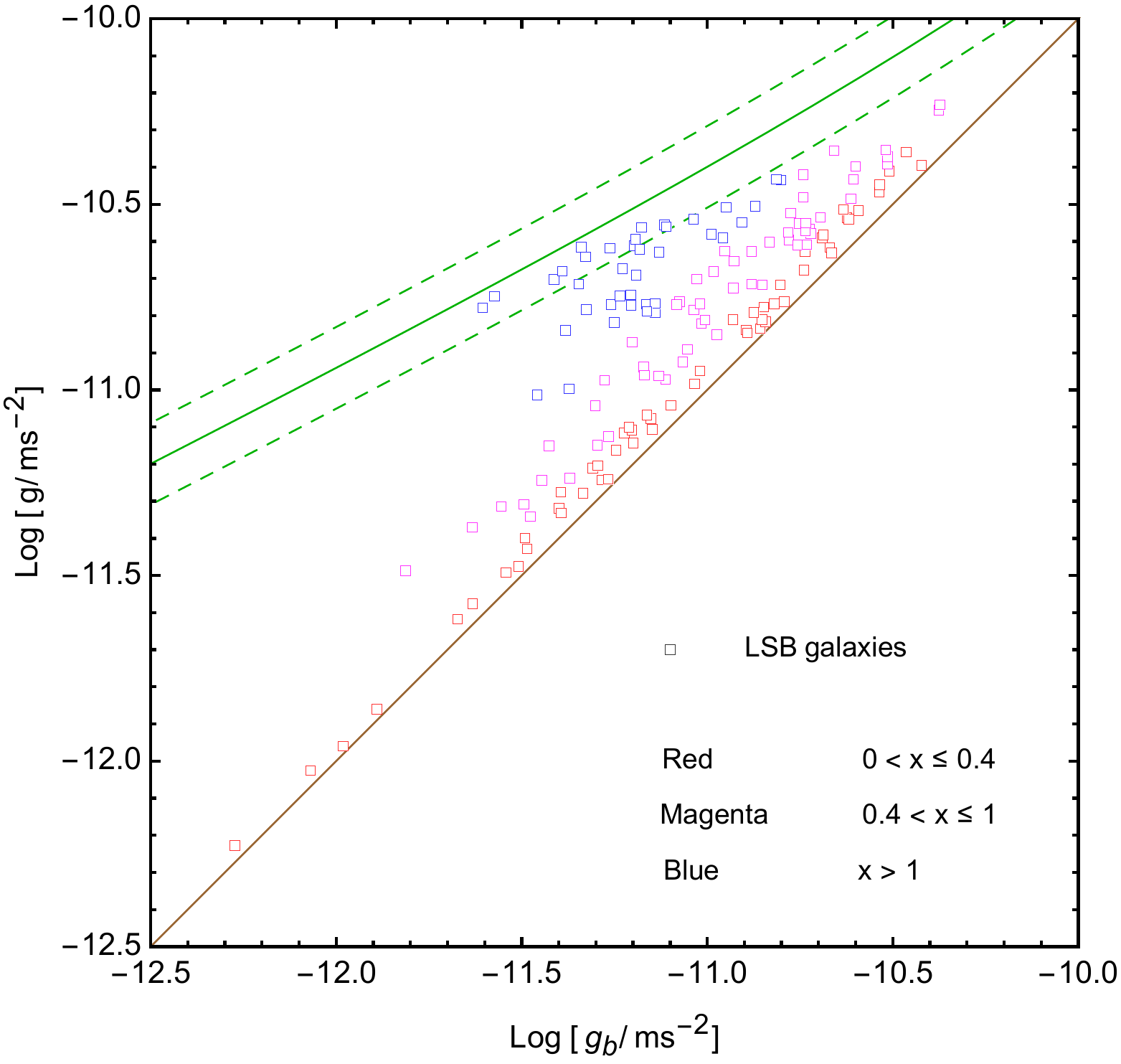}
\end{center} 
\caption{Data resulting from galaxies belonging to the I velocity bin ({\it family}) rotation curve fitted without gas ({\it left panel}) and with gas ({\it right panel}).}
\label{I_velocity_bin_g_gb}
\end{figure}
\noindent 
We have investigated the $g$ - $g_b$ plane by including also the gas component in LSB galaxies when fitting their rotation curves. 
For these galaxies, we 
assumed the contribution of the  gaseous component  by means of the r.h.s. of Eq. \ref{V_stellar} and considering the  mass $M_{HI}$ as 
a free parameter ($M_{HI}$ includes HI + He components).  The results are:
the gas is important only in the first velocity bin; however,
the inner regions are quite dominated by the stellar component and  the gas component is of limited importance.
In Fig. \ref{I_velocity_bin}, we fit the first LSB co-added rotation curve (velocity bin)  without/with the  gas contribution.   
In both cases,  the resulting masses of the  stellar disc and of the DM halo, are similar. In fact, we have:

$M_d = 8.8 \times 10^8 M_{\odot} \;\;$       ;                 $\; r_0 = 10.7 kpc\;\;$ ;    $\;\rho_0= 3.7 \times  10 ^{-3} M_{\odot}/pc^3 \;\;$   ;      $ \; M_{vir} = 1.0 \times  10^{11} M_{\odot} \;\; $. 
\\

\noindent           
While, by considering  
the stellar disc + the DM halo + gaseous disc, we have:

$M_d = 8.0 \times10^8 M_{\odot} \;\;$  ;  $\; r_0 = 10.7 kpc \;\;$  ;      $ \; \rho_0=  3.2 \times  10 ^{-3} M_{\odot}/pc^3 \;\; $   ;        
$ \; M_{vir} = 8.2\times 10^{10} M_{\odot} \;\;$ ;   $ \; M_{HI} = 1.0 \times  10^9  M_{\odot} \;\;$.      
\\     

\noindent
In the above, $M_d, \, r_0, \, \rho_0 ,\, M_{HI}$ (the  quantities obtained by fitting  $V_{co-add}(x, V_{opt})$) are the stellar disc mass, the DM halo core radius, 
the central core mass density, the HI gaseous disc mass (including the correction for helium contribution), respectively.  $M_{vir}$ is the virial mass.

More importantly, we show in Fig. \ref{I_velocity_bin} that the difference in the crucial quantity $f_b$ in the two different cases is small. 
There is only a slightly  
increase at outer galactic radii in the latter case: the resulting data move further  towards the equality line ($g=g_b$), making our results stronger. See Fig. \ref{I_velocity_bin_g_gb}.

\vfill\eject

\section{Fitting uncertainties on $f_b$: the effects on the $g$ - $g_b$ plane}\label{Small errors on $f_b$: small effect on the $g$ - $g_b$ plane}
\begin{figure}[!h]
\begin{center} 
\includegraphics[width=0.6\textwidth,angle=0,clip=true]{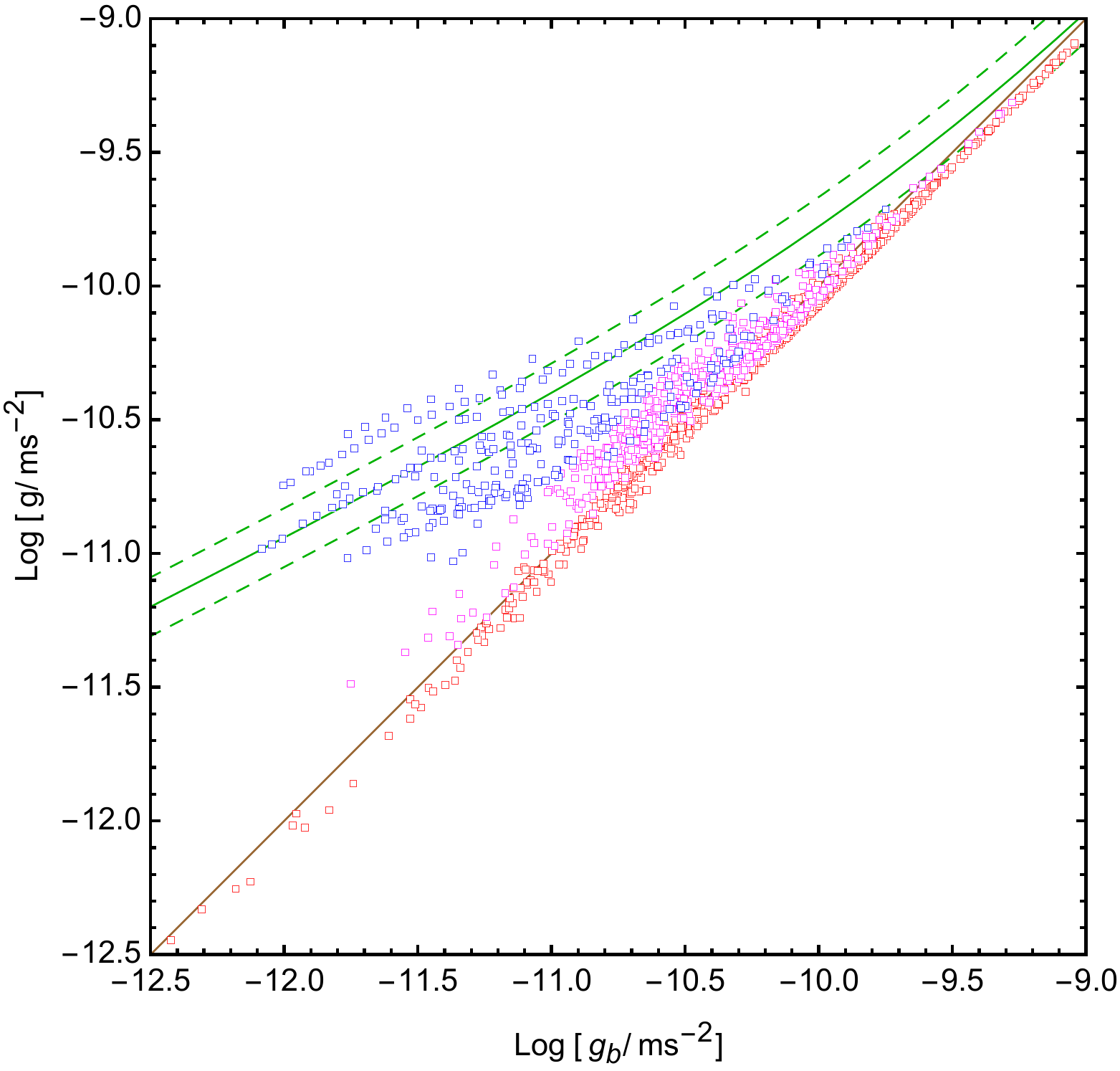} \\\includegraphics[width=0.6\textwidth,angle=0,clip=true]{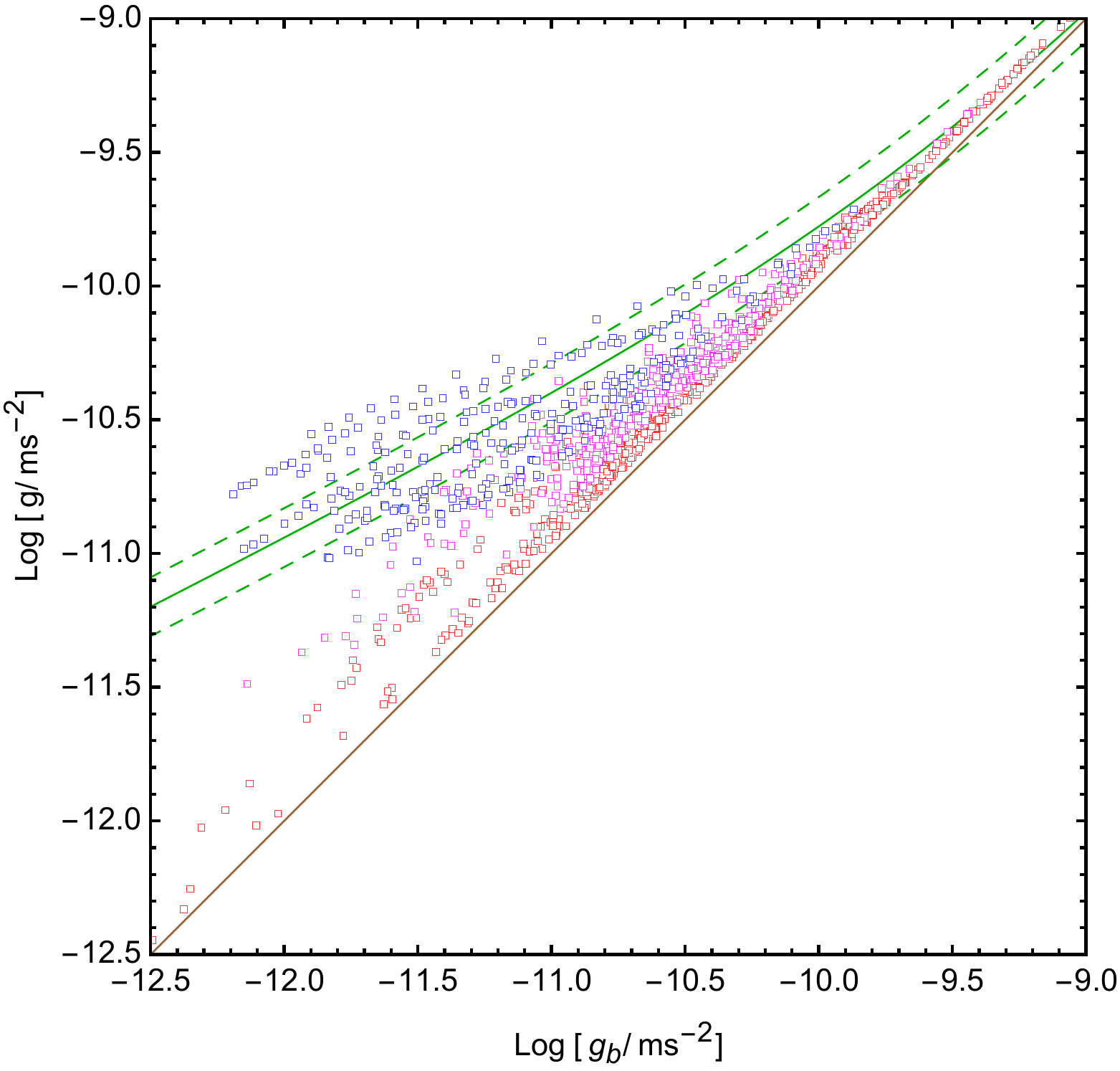}
\end{center}
\caption{The $g$ - $g_b$ relationship by assuming that $f_b$ is $2\sigma$ higher ({\it left panel}) and $2\sigma$ lower than the best value ({\it right panel}).}
\label{g_gb_2sigma}
\end{figure}
\noindent 
The error induced by the kinematical estimation of the stellar mass $M_d$ is very small.
Fig. \ref{g_gb_2sigma} shows the results in the $g$ - $g_b$ plane, 
taking into account a $\pm 2\sigma$ fitting errors on $f_b$ (which is the main source of error). The outcome doesn't change. See Fig. \ref{g_gb_2sigma} and \ref{g_gb_dd_LSB_2D}.

\section{The analysis of the $g$ - $x$ and $g$ - $g_b$ relations in individual galaxies}\label{Single galaxies analysis}
\begin{figure}[!h]
\begin{center} 
\includegraphics[width=0.30\textwidth,angle=0,clip=true]{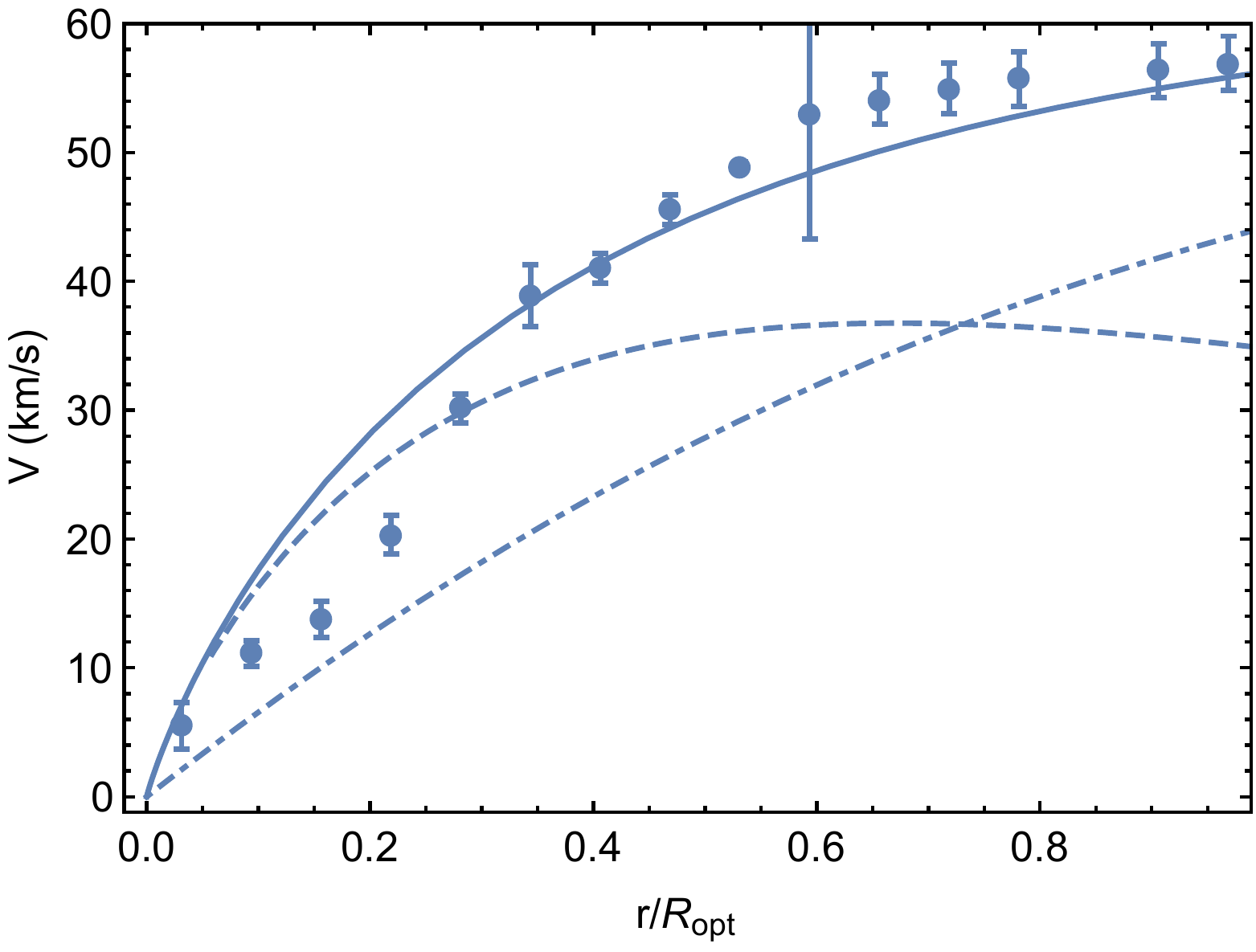}
\includegraphics[width=0.32\textwidth,angle=0,clip=true]{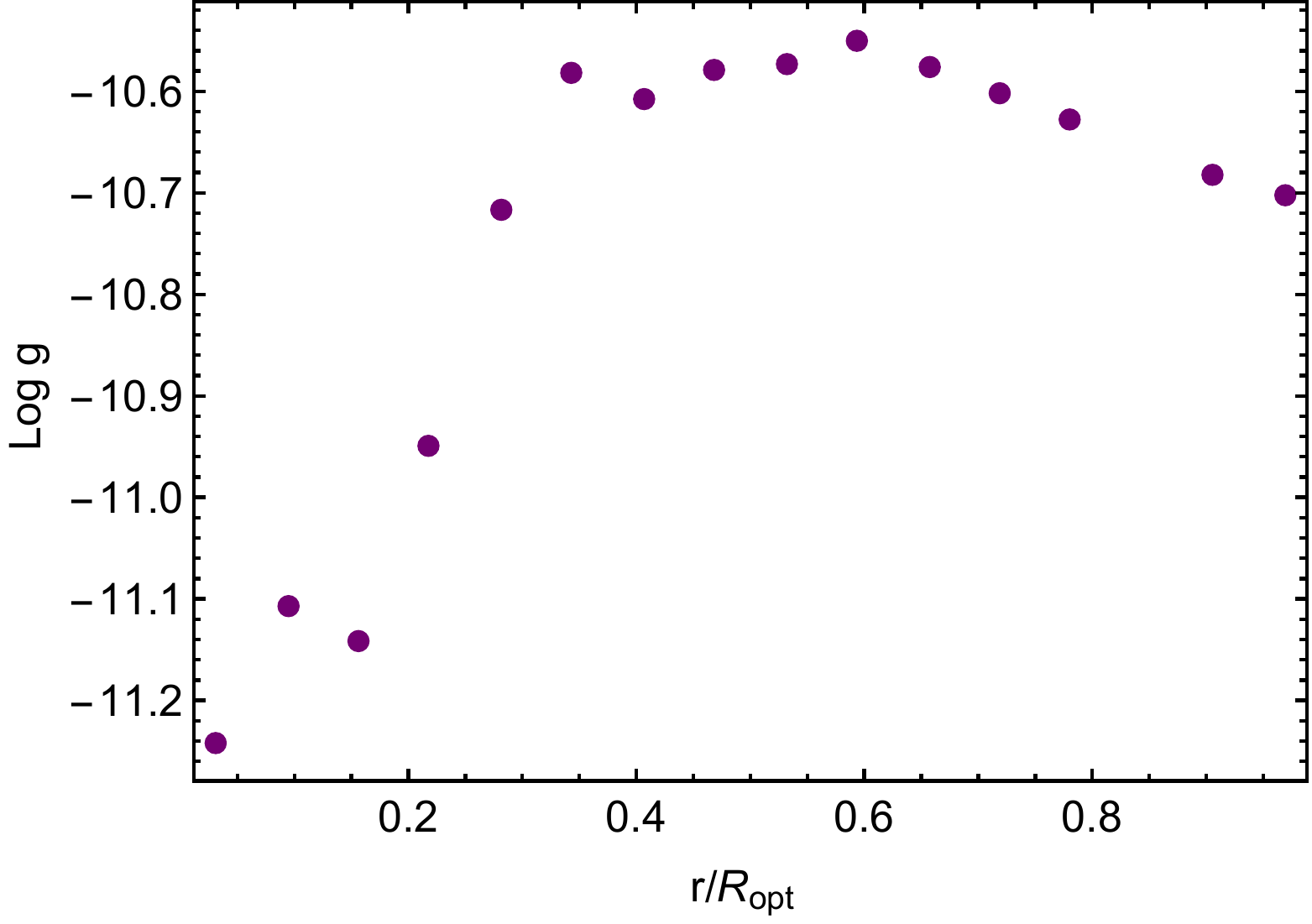}
\includegraphics[width=0.335\textwidth,angle=0,clip=true]{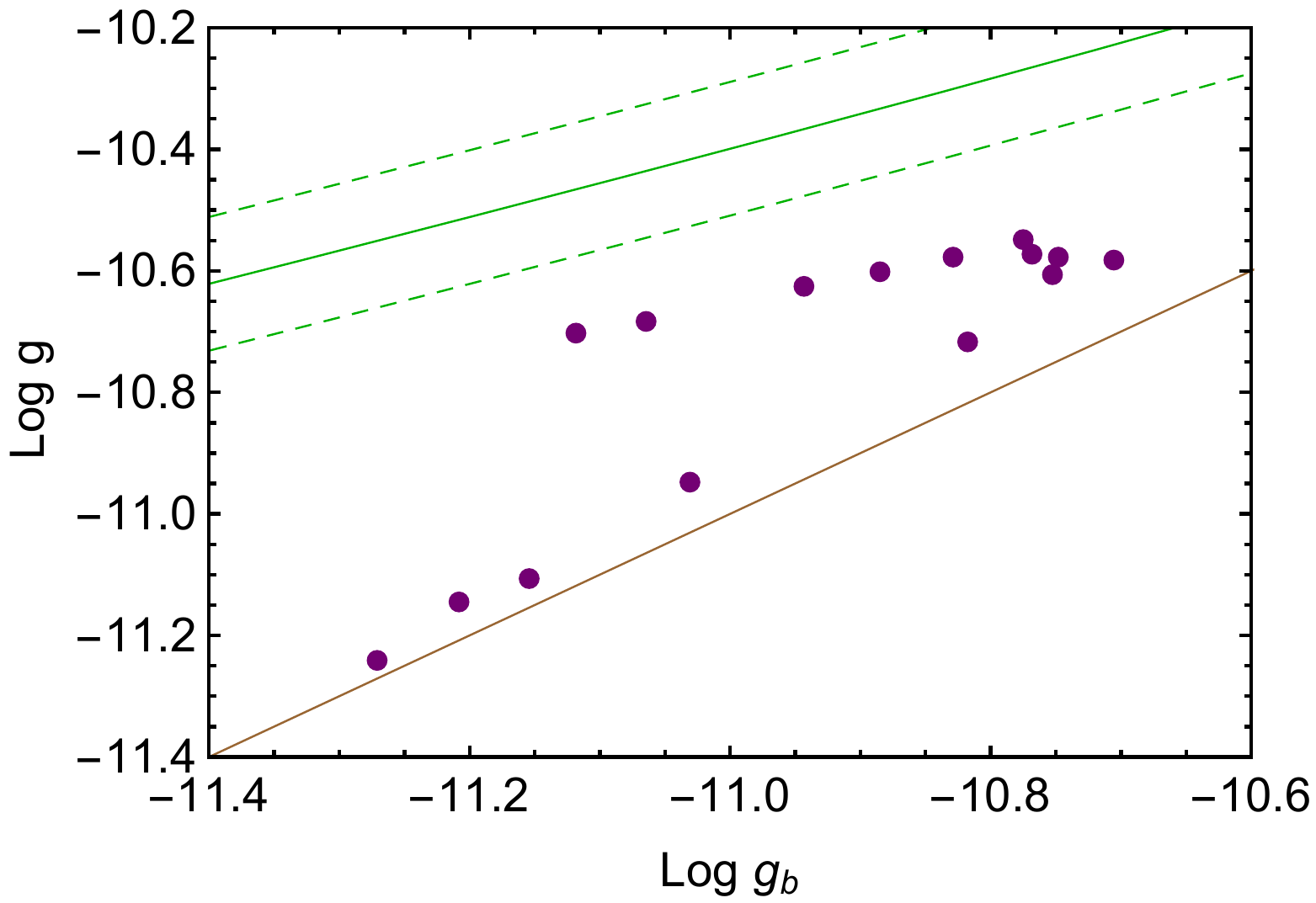} \\
\includegraphics[width=0.30\textwidth,angle=0,clip=true]{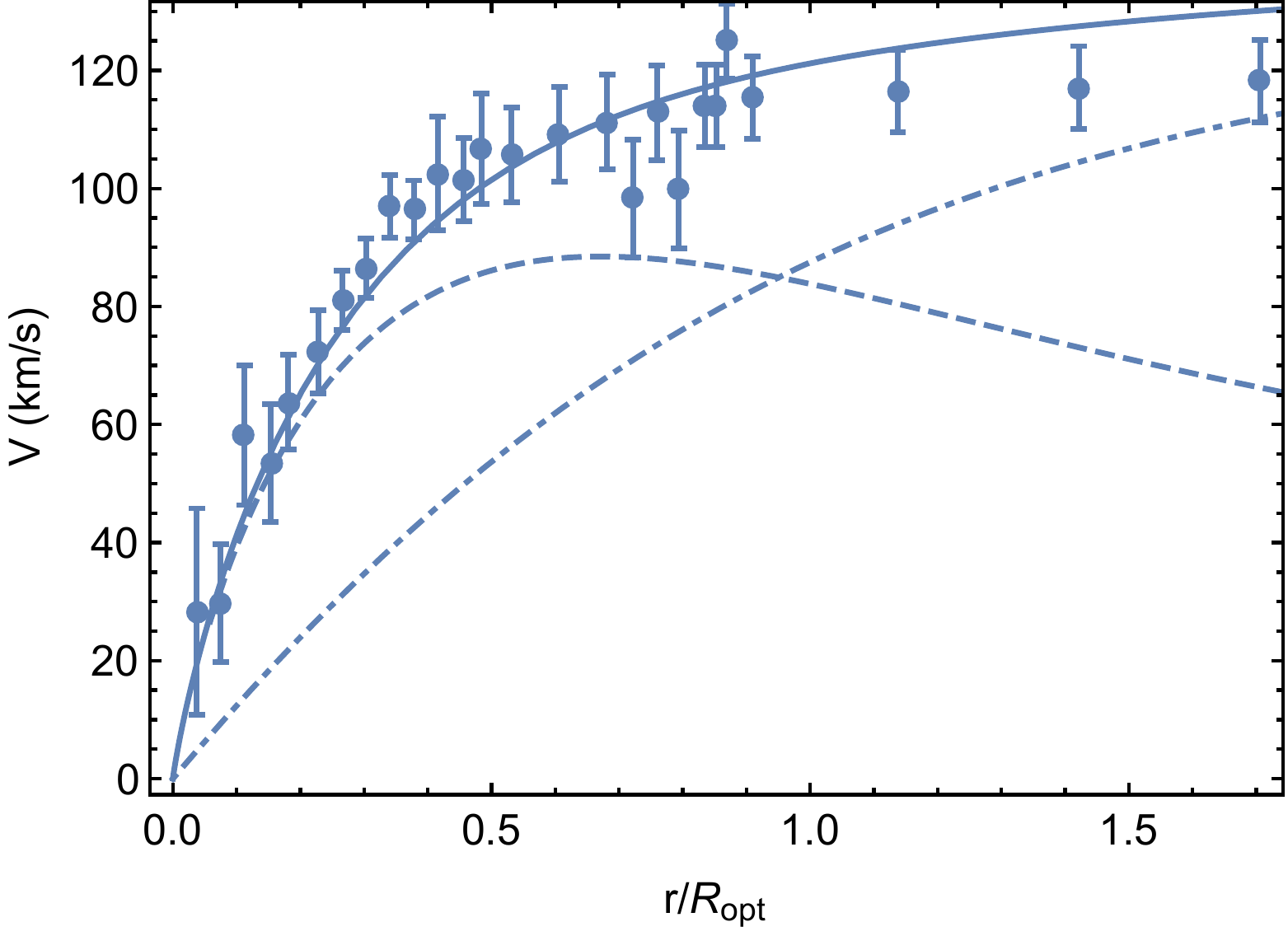}
\includegraphics[width=0.32\textwidth,angle=0,clip=true]{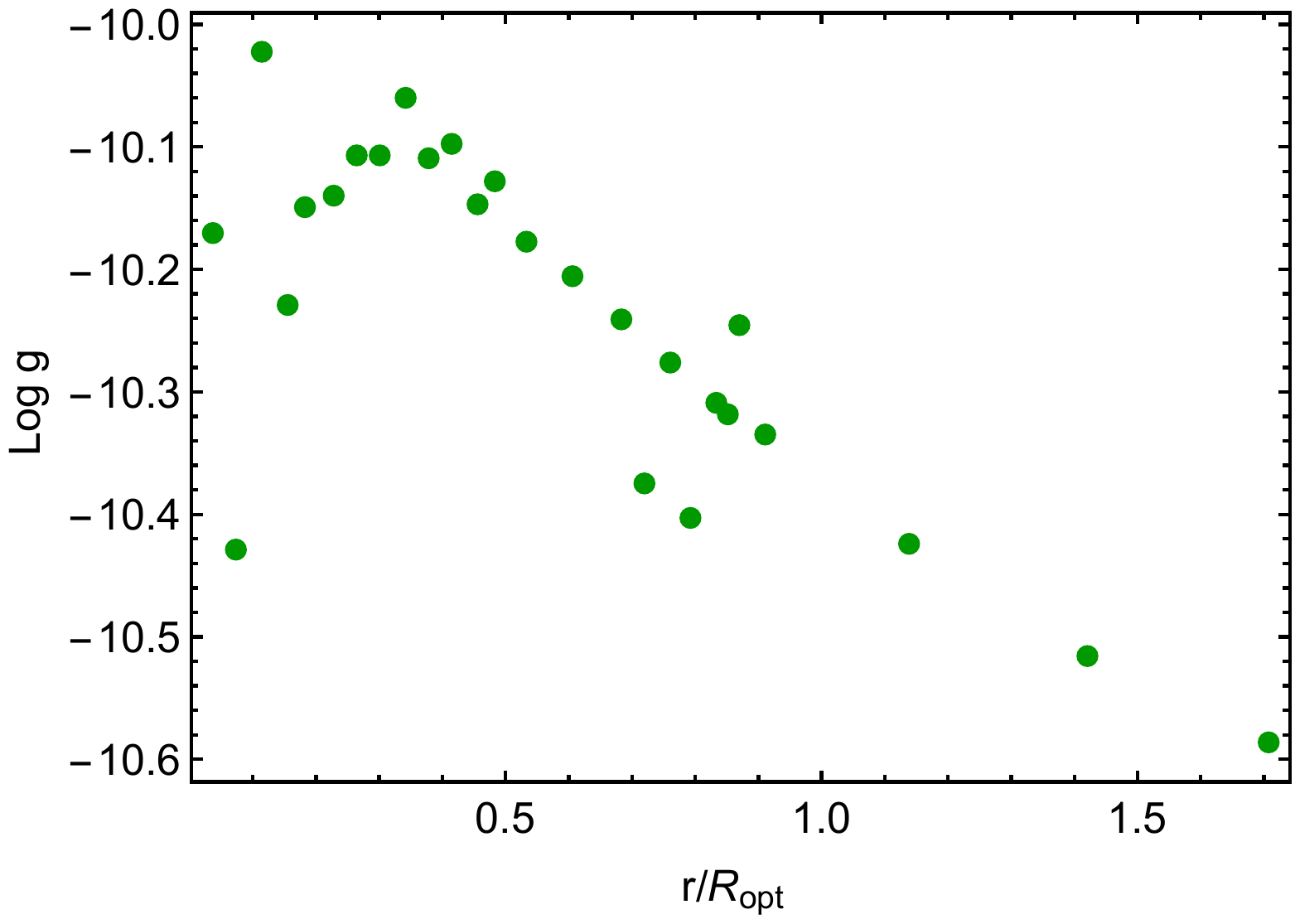}
\includegraphics[width=0.335\textwidth,angle=0,clip=true]{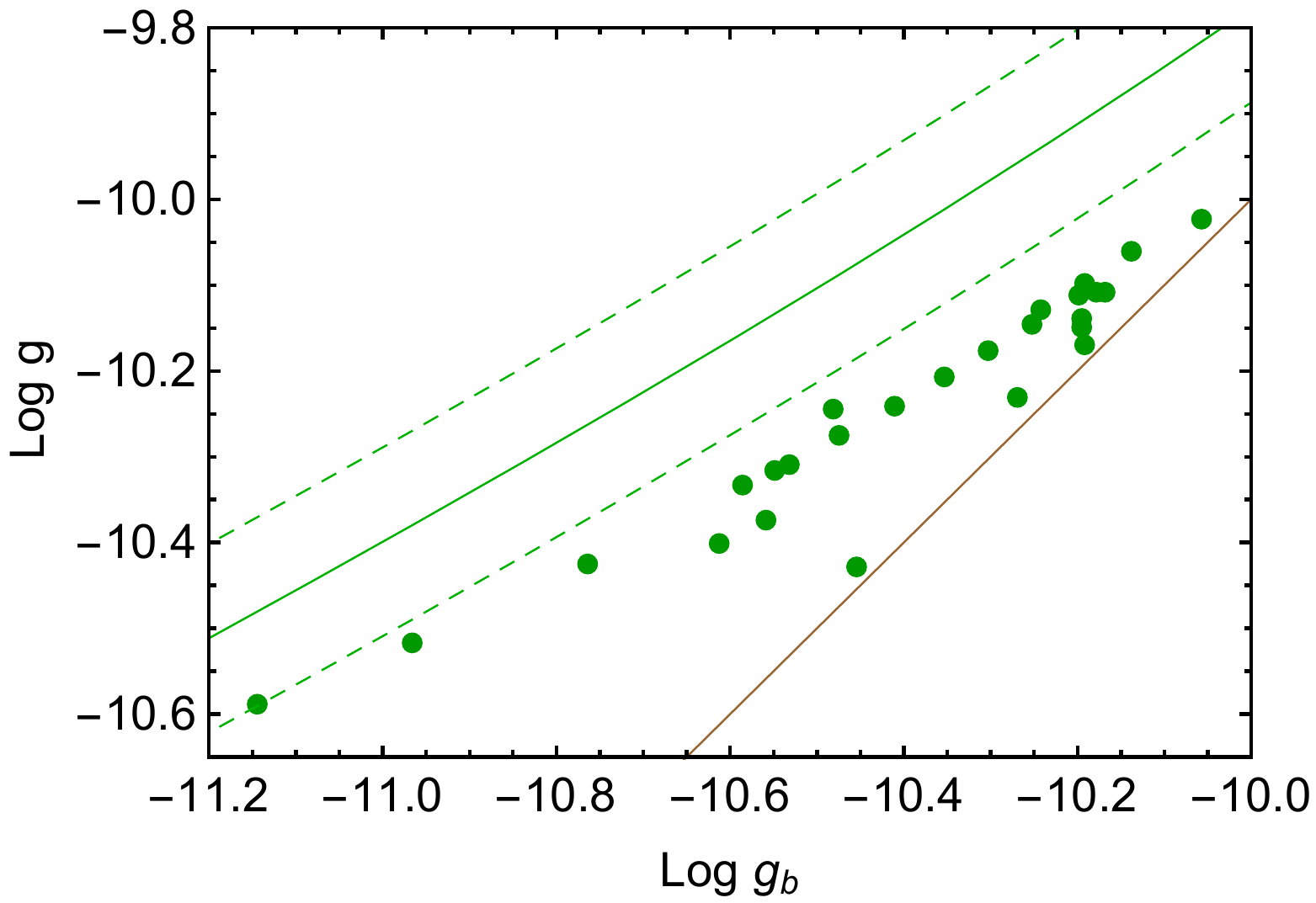} \\
\includegraphics[width=0.30\textwidth,angle=0,clip=true]{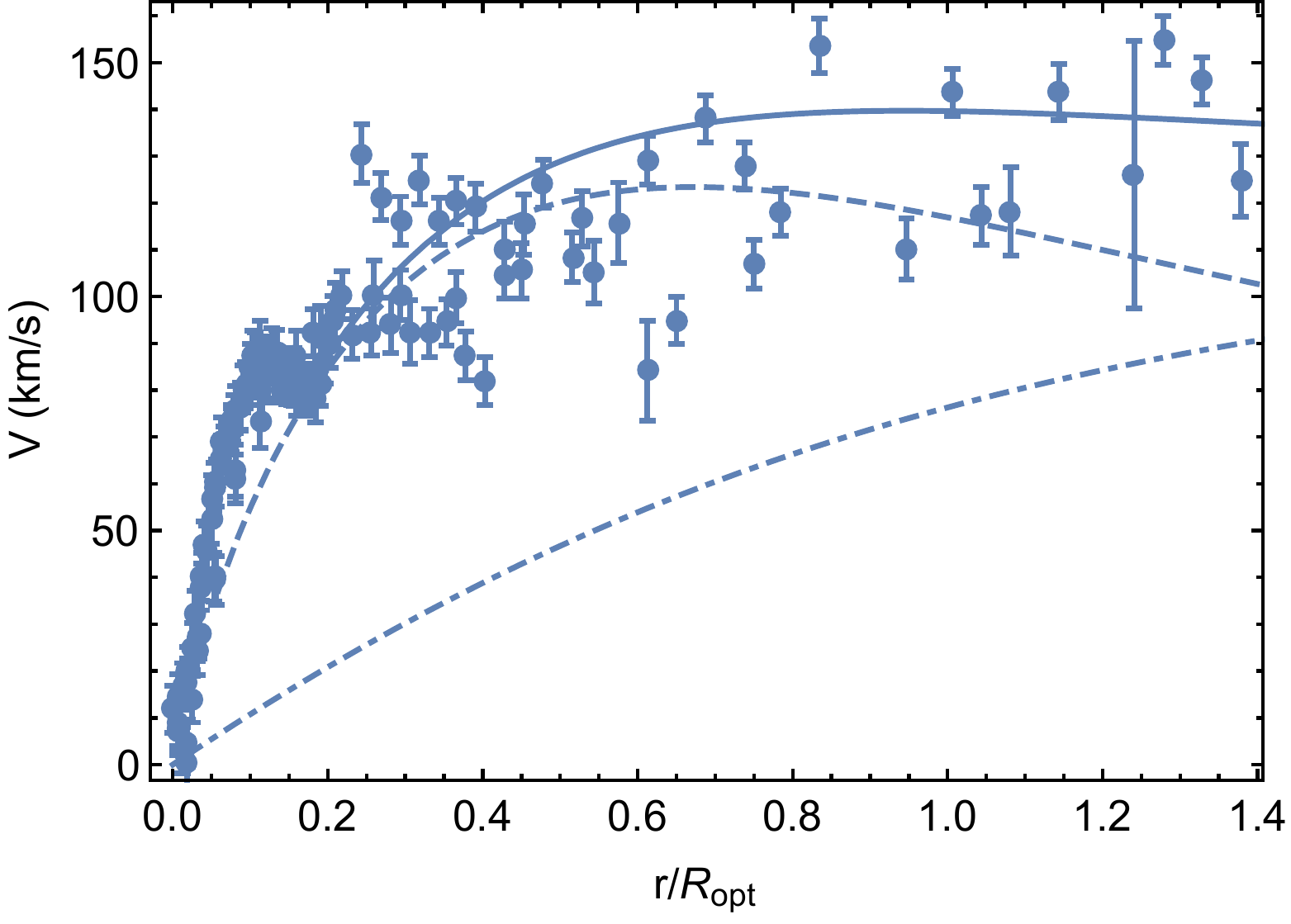}
\includegraphics[width=0.32\textwidth,angle=0,clip=true]{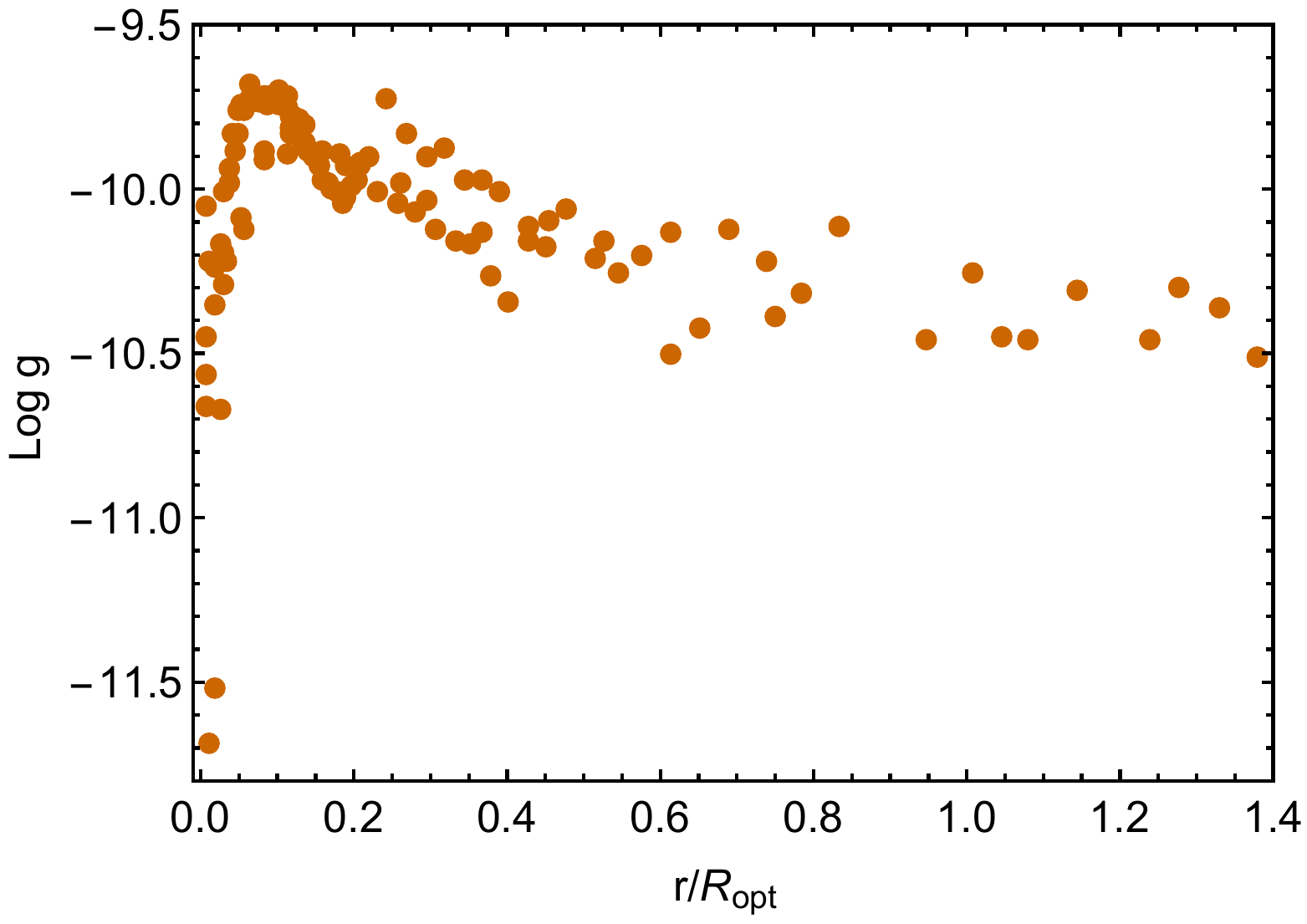}
\includegraphics[width=0.335\textwidth,angle=0,clip=true]{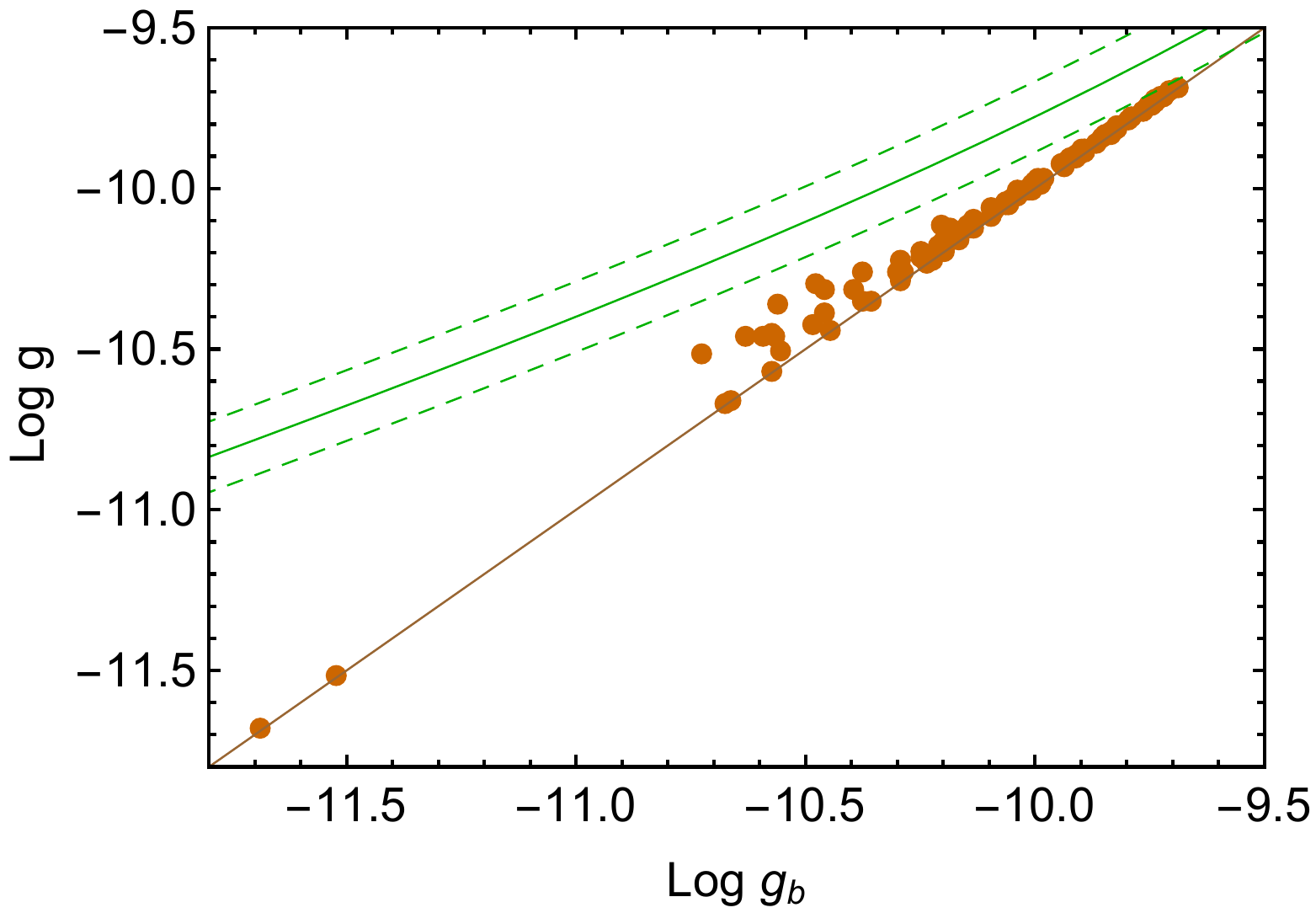}
\end{center} 
\caption{The {\it first, second} and {\it third row} refers respectively to the galaxies: UGC1281, F568V1, ESO234-G013.
 Each row shows the galaxy's rotation curve with fit,  
$g$ vs $r/R_{opt}$ relationship and $g$ vs $g_b$ relationship. 
The {\it green} and the {\it brown lines} are the McGaugh relationship (with its $1\sigma$ uncertainties) and the equality $g= g_b$ relation, 
both independent on the radial coordinate $r/R_{opt}$.}
\label{UGC1281a}
\end{figure}

\noindent 
 It is easier to understand what happens in the $g$ - $g_b$ plane and in the $g$ - $g_b$ - $x$ space by analysing a number of single galaxies.
 Fig. \ref{UGC1281a} shows the  rotation curves, its fits,  
$g$ vs $r/R_{opt}$ relationship and $g$ vs $g_b$ relationship, 
 for three LSBs of different size.  
 The disagreement of present data with McG+16 relationship is evident galaxy by galaxy.
Detailed explanation on this  will appear on Di Paolo et al. (2018) in prep.

\section{The LSB sample}\label{The LSB sample}
\noindent
In Tab. \ref{LSB_sample_Tab}, we report the list of LSB galaxies used in this work and the references of their rotation curves data and other galactic properties.    
\begin{table*}[!t]
\begin{tabular}{p{2.5cm}p{1.3cm}p{1.0cm}p{1.6cm}p{1.5cm}p{4.3cm}}
\hline
Galaxy  &  M  &  Filter                                     &      $R_d$          &     $V_{opt}$      &       Reference \\
              &   mag             &                                                         &       $kpc$          &      $km/s$         &                                    \\      
               (1)       &      (2)       &    (3)                                       &        (4)          &         (5)       &      (6)       \\ 
\hline
NGC 100  &      -19.68  &   I                           &       1.2               &        77.2           &      de Blok \& Bosma, 2002 \\ 
NGC 247	  &      -18.01  &   B                        &         2.9              &        106.6         &     Carignan \& Puche, 1990 \\
NGC 959  &	-18.53  &  V                           &         0.93            &      75.3               &     Kuzio de Naray et al. 2008  \\
NGC 2552  &	-18.99  &  I                            &        1.6              &       104.9             &     Kuzio de Naray et al. 2008 \\
NGC 2552   &	-18.99  &  I                             &        1.6              &        111.0                &     de Blok \& Bosma, 2002 \\
NGC 2552  &	-18.99  &  I                                       &         1.6             &        92.6            &     Swaters et al. 2003 \\
NGC 2552  &	-18.1  &     R                 &          1.6          &        92.5             &     van den Bosch \& Swaters, 2001  \\
NGC 3274   &	-16.7  &    R                             &          0.5            &      79.7              &       de Blok \& Bosma, 2002    \\
NGC 3274  &	-16.6   &  R                                   &   0.45                 &      63.2             &        Swaters et al. 2003        \\
NGC 3347B  &	-21.76  &  I                           &       8.1              &      167.0                &      Palunas \& Williams, 2000   \\
NGC 4395  &     -18.1    &  R                      &        2.3               &      82.0             &      de Blok \& Bosma, 2002  \\
NGC 4395  &  -18.14  &  R                 &        2.6             &     82.6                &    van den Bosch \& Swaters, 2001 \\
NGC 4455  &  -16.9    &  R                         &        0.7              &     45.6                &    de Blok \& Bosma, 2002  \\
NGC 4455  &  -16.88  &  R                                &       0.9               &       61.9             &    Marchesini et al. 2002     \\
NGC 4455  &  -16.88  &  R                 &        0.9              &      51.5               &    van den Bosch \& Swaters, 2001 \\
NGC 5023  &  -19.18  &  I                               &         0.8               &     78.4               &    de Blok \& Bosma, 2002   \\
NGC 5204  &  -17.3  &     R                                     &        0.66              &    75.2                &    Swaters et al. 2003   \\
NGC 5204  &  -17.28  &  R                   &        0.66               &     71.0              &    van den Bosch \& Swaters, 2001  \\
NGC 7589  &  -21.9  &    R                                    &         13            &      224.0                  &   Pickering et al. 1997   \\
UGC 628    &  -19.2  &    R                           &          4.7           &      130.0                   &    de Blok \& Bosma, 2002   \\
UGC 634  &     -17.7  &  B                                       &             3.1        &        95.1              &    van Zee et al. 1997 \\
UGC 731  &    -16.6  &  R                            &           1.7           &          73.1            &   de Blok \& Bosma, 2002  \\
UGC 731  &    -16.6  &  R                                        &            1.6        &       73.5               &    Swaters et al. 2003    \\
UGC 731  &    -16.63  &  R                &           1.6        &        73.5               &    van den Bosch \& Swaters, 2001  \\
UGC 1230  &  -19.1    &  R                           &              4.5        &      104.5              &    de Blok \& Bosma, 2002   \\
UGC 1230  &  -17.16  &  NUV                        &            4.4         &      89.7                &     van der Hulst et al. 1993   \\
UGC 1281  &  -16.2  &    R                        &            1.7          &        45.8                &   Kuzio de Naray et al. 2006   \\
UGC 1281  &  -16.2   &  R                            &             1.7         &     56.9                 &    de Blok \& Bosma, 2002  \\
UGC 1551  &  -19.7  &    B                         &            2.5        &    55.8                    &     Kuzio de Naray et al. 2008\\
UGC 2684  &  -13.7  &    B                                 &            0.8          &      36.7                &   van Zee et al. 1997   \\
UGC 2936  &  -21.1  &    R                             &             8.4         &       255.0              &   Pickering et al. 1999   \\
UGC 3137  &  -18.7  &    R                         &            2.0              &     97.7                 &   de Blok \& Bosma, 2002 \\
UGC 3174  &  -15.7  &    B                                 &            1.0               &     51.7                &  van Zee et al. 1997  \\
UGC 3371  &  -17.7  &    R                    &         3.1             &        84.7               &  de Blok \& Bosma, 2002 \\
UGC 3371  &  -17.74  &  R                 &         3.1           &      85.1                &  van den Bosch \& Swaters, 2001 \\
UGC 4115  &       -15.21   &  V                                &          0.4           &    24.2                    & McGaugh et al. 2001 \\
UGC 4278  &  -17.7   &  R                           &            2.3          &   92.6                   & de Blok \& Bosma, 2002 \\
UGC 5005  &  -17.8  &    B                        &           4.4            &    95.5                 & de Blok \& McGaugh, 1997 \\
UGC 5272  &  -14.7  &    B                      &            1.2            &     51.2                   & Kuzio de Naray et al. 2008 \\
UGC 5272  &  -14.7  &    B                       &           1.2            &    46.4                   & de Blok \& Bosma, 2002  \\
UGC 5716  &  -16.3   &   B                                    &             2.0            &       66.4                & van Zee et al. 1997  \\
UGC 5750  &  -19.5  &    R                           &           5.6            &    58.5                    & Kuzio de Naray et al. 2006 \\
UGC 5750  &  -19.5  &    R                          &       5.6                &   78.5                  &   de Blok \& Bosma, 2002 \\
UGC 5999  &  -12.42  &  R                            &          4.4              &    153.0                &  van der Hulst et al. 1993 \\
UGC 7178  &  -16.6  &    B                              &        2.3               &       69.9                 & van Zee et al. 1997 \\
UGC 8837  &  -15.7  &    R                              &         1.2              &      49.6                &   de Blok \& Bosma, 2002  \\
UGC 9211    &  -16.21  &  R                &         1.3           &      61.9                   &  van den Bosch \& Swaters, 2001 \\
UGC 11454  &  -22.03  &  R                             &   4.5                   &   150.3                    &   McGaugh et al. 2001  \\
UGC 11557  &  -19.7  &    R                                   &     3.1                  &     83.7                 &   Swaters et al. 2003   \\
UGC 11583  &  -15.48  &  R                                   &       0.31              &    27.9                  &   McGaugh et al. 2001 \\
\hline 
\end{tabular}
\caption{LSB sample. Columns: (1) galaxy name; (2) magnitude, given for further information of the galaxies; (3) filter; (4) stellar disc scale length $R_d$; 
(5) optical velocity $V_{opt}$; (6) reference. 
Note that some galaxies have multiple rotation curve data, that we have homogenised.}
\label{LSB_sample_Tab}
\end{table*}
\begin{table*}[!t]
\begin{tabular}{p{2.5cm}p{1.3cm}p{1.0cm}p{1.6cm}p{1.5cm}p{4.3cm}}
\hline
Galaxy  &  M  &  Filter                                       &      $R_d$          &     $V_{opt}$      &       Reference\\
              &       mag         &                                                              &       $kpc$          &      $km/s$         &                                    \\    
               (1)       &      (2)       &    (3)                                       &        (4)          &         (5)       &      (6)       \\   
\hline
UGC 11616  &  -21.58  &  R                                 &        4.9                 &    133.2          &  McGaugh et al. 2001 \\
UGC 11648  &  -22.95  &  KS                              &        3.8                 &    142.2        &   McGaugh et al. 2001 \\
UGC 11748  &  -23.02  &  R                                &       3.1               &   240.7          &  McGaugh et al. 2001 \\
UGC 11819  &  -20.62  &  R                                &        5.3                 &    154.6          &   McGaugh et al. 2001  \\
ESO 186-G055  &    -20.62  &  R                           &        3.6               &     133.2        &  Pizzella et al., 2008  \\
ESO 206-G014  &    -20.32  &  R                          &        5.2              &      91.3            & Pizzella et al. 2008  \\
ESO 215-G039  &    -21.72  &  I                  &         4.2                &     142.9         & Palunas \& Williams, 2000  \\
ESO 234-G013  &    -21.66  &  I                            &          3.7              &    139.4         &   Pizzella et al. 2008  \\
ESO 268-G044  &    -21.19  &  I                 &         1.9               &    175.6         &   Palunas \& Williams, 2000 \\
ESO 322-G019  &    -20.41  &  B                 &          2.5              &      100.7        &   Palunas \& Williams, 2000   \\
ESO 323-G042  &    -21.56  &  I            &          4.4              &     138.7          &   Palunas \& Williams, 2000  \\
ESO 323-G073  &    -21.81  &  I                   &         2.1                &      165.3       &   Palunas \& Williams, 2000 \\
ESO 374-G003  &    -21.36  &  I                  &         4.2                  &    118.3       &  Palunas \& Williams, 2000 \\
ESO 382-G006  &    -17.03  &  R                  &    2.3                    &   160.0             &   Palunas \& Williams, 2000  \\
ESO 400-G037  &    -20.96  &  I                              &         4.1             &   69.9             &    Pizzella et al. 2008  \\
ESO 444-G021  &    -19.9  &  B                      &       6.4               &     107.4      &   Palunas \& Williams, 2000 \\ 
ESO 444-G047  &    -21.11  &  I                      &       2.7               &     148.4       &   Palunas \& Williams, 2000  \\
ESO 488-G049  &    -17.94  &  B                                &       4.4               &      95.3          &    Pizzella et al. 2008 \\
ESO 509-G091  &    -21.01  &  I                      &         3.7              &       146.8      &  Palunas \& Williams, 2000  \\
ESO 534-G020  &    -21.96  &  R                              &        17              &      216.6       &   Pizzella et al. 2008   \\
F561-1  &  -17.8  &    B                                            &      3.6                   &       50.8     &      de Blok et al. 1996  \\
F563-V1  &  -16.3  &    B                                           &       2.4                   &     27.3       &    de Blok et al. 1996   \\
F563-V2  &  -18.2  &    B                              &        2.1                 &      98.8           &   Kuzio de Naray et al. 2006  \\
F563-V2  &  -17.6  &    B                                          &        2.1                &       98.0        &  de Blok et al. 1996 \\
F565-V2  &  -14.8  &    B                                            &       2.0                     &          45.2      &    de Blok et al. 1996 \\
F568-1  &  -18.1  &    B                                                &       5.3                &      130.1           &   Swaters et al. 2000  \\
F568-3  &  -19.14  &  I                                      &     4.0                     &       102.6          &   Kuzio de Naray et al. 2006  \\
F568-3  &  -18.3    &  B                                            &      4.0                   &       97.9             &  McGaugh et al. 2001 \\
F568-3  &  -18.3  &    B                                            &          4.0                 &      101.1           &  Swaters et al. 2000  \\
F568-6  &  -23.6  &    R                                            &             18         &    297.0           &    Pickering et al. 1997 \\
F568-V1  &  -17.9  &    B                                         &        3.2                &    115.8         &   Swaters et al. 2000  \\
F571-8  &  -17.6  &    B                                            &     5.2                 &      139.4          &   Marchesini et al. 2002 \\
F571-8  &  -17.6  &    B                                            &         5.2             &       140.1         &  McGaugh et al. 2001  \\
F571-V1  &  -11.47  &  I                                             &        3.2               &    72.44            &  de Blok et al. 1996    \\
F574-1  &  -18.4    &  B                                               &        4.3               &   102.3           &  Swaters et al. 2000   \\
F574-2  &  -17  &     B                                                   &        4.5              &   40.0               &   de Blok et al. 1996\\
F579-V1  &  -18.8  &    B                                         &        5.1               &  111.5            &  McGaugh et al. 2001 \\
F583-1  &  -16.5  &    B                                &     1.6                 &     68.2          & Kuzio de Naray et al. 2008  \\
F583-1  &  -17.06  &  R                                          &      1.6                &      65.2          & Marchesini et al. 2002  \\
F583-1  &  -16.5    &  B                                              &    1.6                  &       61.3           &    McGaugh et al. 2001   \\
F583-1  &  -15.9  &    B                                                &    1.6                &      53.3            &   de Blok et al. 1996  \\
F583-4  &  -16.9  &    B                                    &    2.7                 &      83.9              & Kuzio de Naray et al. 2006 \\
F583-4  &  -16.9  &    B                                             &     2.7                 &       69.0               &   McGaugh et al. 2001  \\
F730-V1  &  -20.27  &  R                                          &      5.8              &     141.6             & McGaugh et al. 2001 \\
PGC 37759  &  -21.88  &  Z                                           &     6.6             &      139.4             &    Morelli et al. 2012  \\
\hline 
\end{tabular}
\caption{It continues from Tab. \ref{LSB_sample_Tab}.}
\label{LSB_Tab_3}
\end{table*}
\\
\\
\\
\\

\vfill\eject

\bibliographystyle{aasjournal}
\bibliography{Bibliografia}

\end{document}